\documentclass[useAMS, usenatbib]{mn2e}
\usepackage{graphicx}

\title[Disk models within $\Lambda$CDM haloes]{Idealized models for galactic disk formation and evolution in `realistic' 
       $\Lambda$CDM haloes}
\author[M. Aumer and S.D.M. White]
{Michael Aumer$^{1}$ \thanks{E-mail:maumer@mpa-garching.mpg.de (MA)} and Simon D.M. White$^{1}$\\
$^{1}$Max-Planck-Institut f\"ur Astrophysik, Karl-Schwarzschild-Str. 1, 85748 Garching, Germany}

\begin{document}

\date{draft version}

\pagerange{\pageref{firstpage}--\pageref{lastpage}} \pubyear{2012}

\maketitle

\label{firstpage}

\begin{abstract}
We study the dynamics of galactic disk formation and evolution in `realistic' $\Lambda$ cold dark matter haloes
with idealized baryonic initial conditions. We add rotating spheres of hot gas at $z=1.3$ to two fully
cosmological dark-matter-only halo (re)simulations. The gas cools according to an artificial and adjustable cooling
function to form a rotationally supported galaxy. The simulations evolve in the full cosmological context until z=0. 
We vary the angular momentum and density profiles of the initial gas sphere, the cooling time and the orientation
of the angular momentum vector to study the effects on the formation and evolution of the disk. The final disks
show exponential radial and (double)-exponential vertical stellar density profiles and stellar velocity dispersions
that increase with age of the stars, as in real disk galaxies. The slower the cooling/accretion processes, the
higher the kinematic disk-to-bulge (D/B) ratio of the resulting system. We find that the initial orientation
of the baryonic angular momentum with respect to the halo has a major effect on the resulting D/B. The most stable
systems result from orientations parallel to the halo minor axis. Despite the spherical and coherently rotating 
initial gas distribution, the orientation of the central disk and of the outer gas components and the relative
angle between the components can all change by more than 90 degrees over several billion years. Initial orientations
perpendicular to the major axis tend to align with the minor axis during their evolution, but the \textit{sign of the
spin} can have a strong effect. Disks can form from initial conditions oriented parallel to the major axis, but
there is often strong misalignment between inner and outer material. The more the orientation of the baryonic angular
momentum changes during the evolution, the lower the final D/B.  The behaviour varies strongly from halo to halo.
Even our very simple initial conditions can lead to strong bars, dominant bulges, massive, misaligned rings and
counter-rotating components. We discuss how our results may relate to the failure or success of fully cosmological
disk formation simulations.
\end{abstract}


\section{Introduction}

The majority of galaxies in the local universe with masses similar to that of the Milky Way are disk-dominated (e.g. \citealp{delgado}).
In the standard paradigm, these disks formed through cooling and condensation of gas
within cold dark matter haloes \citep{whiterees, fall}. Disk size is a consequence of the angular momentum of the gas, which had
previously been acquired through tidal torques from neighboring structures (e.g. \citealp{peebles, white}).

Numerical hydrodynamical simulations of this formation scenario have been carried out in great number
(e.g. \citealp{navarro94, navarro2, abadi, governato, aquila}). Despite recent progress (e.g. \citealp{eris, agertz, sales}), these simulations
have in general suffered from a range of problems, including angular momentum loss leading to overly small disks \citep{navarro1},
production of overly massive galaxies \citep{guo}, and too much early and too little late
star formation \citep{aquila}. Even if extended disks form, they are often destroyed by infalling satellites \citep{toth}
or the accretion of misaligned gas \citep{aquila, sales}. Moreover, different numerical schemes can yield very different galaxies
for the same initial conditions \citep{cs2011,keres}. In particular, the angular momentum content and thus the
structure of the disks depends strongly on the numerical method applied \citep{torrey}.
Overall, the galaxy population predicted by cosmological gas dynamical simulations disagrees in major ways with observations.
Since detailed population properties are reproduced quite well in the standard $\Lambda$CDM paradigm by simple semi-analytic simulations
(e.g. \citealp{guo2}), it has been argued that the relevant star formation and feedback processes are still
inadequately represented in hydrodynamical simulations.
Such semi-analytic models are however not capable of properly capturing the complex dynamical interactions
between gas, stars and dark matter found to be important in the above-mentioned simulations.

In contrast, there is general agreement on the formation and structure of dark matter haloes in $\Lambda$CDM
(e.g. \citealp{aquarius}), and on the observational side, a detailed picture of the structure of disk galaxies has been assembled
over the last decades (see \citealp{freeman} for a recent review).
In this study, we therefore study whether  the detailed output of simulations of dark matter halo formation
and our detailed knowledge of disk galaxies can be brought into agreement if the fully cosmological treatment of baryon physics
is replaced by idealized models.

Previous attempts in this direction include the models of \citet{weil}, who showed that the prevention of radiative gas cooling
in cosmological hydrodynamical simulations until $z=1$ allows the formation of a disk galaxy population with realistic 
angular momentum content. \citet{kaufmann, kaufmann2} simulated disk formation by allowing
a rotating gas distribution to cool inside idealized (mostly spherical) equilibrium haloes.
This enabled them to examine numerical effects as well as some physical processes related to disk simulations.

$\Lambda$CDM haloes, however, show substructure, are triaxial and are continuously
accreting \citep{frenk}. The impact of halo shape is still an open issue.
Disk galaxies in cosmological hydrodynamical simulations are typically aligned with the halo minor axis \citep{bailin}.
Moreover, it has been argued that the interaction of the forming disk galaxy with the dark matter
tends to make the haloes axisymmetric (e.g. \citealp{berentzen, kazantzidis}). 
\citet{berentzen} introduced a disk galaxy within a cosmological triaxial
halo by gradually adding stellar disk particles according to an axisymmetric analytical disk model with growing mass and size,
to the simulation. They were thus able to analyze the interplay between a galactic disk and a triaxial halo.
A similar study is currently being undertaken for the same dark matter haloes we use here by DeBuhr et al. (in prep.).

For our study, we perform a set of controlled numerical experiments using simplified prescriptions for 
the physics of gas accretion and star formation in order to gain insight into the dynamical processes that 
affect the formation of galaxy disks within $\Lambda$CDM haloes . We use fully cosmological, triaxial haloes as initial conditions, but 
simulate disk formation and evolution by the cooling of a rotating gas sphere starting at redshift $z=1.3$. 
Our idealized treatment allows us to study how 
the formation of a galactic disk is affected by the rapidity of gas cooling and by the angular momentum and mass of the gas and by 
the orientation of its angular momentum with respect to the principal axes of the dark halo. 
We compare our results to the direct output of fully cosmological hydrodynamical simulations 
carried out for the same haloes, so that we can better understand why they fail to produce substantial disks.

In Section 2 we describe the setup and numerical methods of this work.
In Section 3 we describe the formation, evolution and structure of one particular disk model.
In Section 4 we analyze the dependence of the disk growth on our model parameters.
In Section 5 we discuss the influence of initial spin orientation and the stability of the orientation of the resulting disks.
In Section 6 we compare our models to the fully cosmological galaxy formation simulations of \citet{aquila} (CS09 hereafter).
Finally, in Section 7 we summarize and conclude.

\section{Simulation setup}

To mimic the formation of a galactic disks within fully-cosmological
dark-matter-only simulations of haloes expected to host Milky-Way type galaxies,
we make use of the Aquarius simulations \citep{aquarius}, a suite of high resolution zoom-in resimulations of six dark matter haloes. 
These haloes were chosen from a simulation of a cosmological box with a side-length of $137\; \rm {Mpc}$, and were simulated from
$z=127$ assuming a $\Lambda$CDM universe with the following parameters: $\Omega_{\Lambda}=0.75$, $\Omega_m=0.25$,
$\Omega_b=0.04$, $\sigma_8=0.9$ and $H_0=73\;\rm{kms}^{-1}\rm{Mpc}^{-1}$. For details we refer to \citet{aquarius}.

We select two haloes, named A and C, both at resolution level 5, which corresponds to dark matter particle masses of
$m_{\rm{dm}}\approx 3\times10^6 M_{\odot}$ and a gravitational softening length of $\epsilon_{\rm{dm}}=685\;\rm{pc}$. 
These haloes have (dark matter only) $z=0$ virial masses of $M_{200}=1.853$ and $1.793\times 10^{12} M_{\odot}$
and have been studied in the fully cosmological galaxy formation simulations of CS09.
We selected halo A as it has a very quiescent merger history after redshift $z\sim1$ and has been identified
as a prime candidate for hosting a disk galaxy by semi-analytical modeling. However, CS09 did not find
a significant disk component in their $z=0$ simulations. Halo C, in contrast, showed the highest (yet still unrealistically low)
disk-to-bulge ratio, $ D/B  \sim 1/4$, of all 8 haloes studied in this work.
Halo C was also studied in the \textit{Aquila Comparison Project} \citep{cs2011}, which compared the results
of various cosmological gas-dynamical codes.

For our simulations we apply the TreeSPH-code GADGET-3, last described in \citet{gadget}, 
an extended version of which was used by CS09.

As initial conditions for our numerical experiments we use the simulation outputs of \citet{aquarius} at $z=1.3$.
We choose this time, as it falls after the epoch of major mergers.
Disk galaxies have very likely been in place and continuously forming since before $z=1.3$.
For example, age estimates for the solar neighbourhood exceed $10 \;\rm {Gyr}$ (see \citealp{ab} and references therein),
making our choice appear problematic. It has been argued that the ages of thin disk stars are consistent with formation
after $z \sim 1.5$ (e.g. \citealp{reddy}), but it is not clear, if a two-component-division of the Milky Way disk
into thin and thick components is sensible (e.g. \citealp{sb}).
Our goal is not however to model the entire formation and evolution of disk galaxies,
but to study the ability of realistic dark matter haloes with evolving substructure to host thin disk galaxies over
cosmological times.

To simulate the formation of a disk galaxy within the dark matter halo, we insert a rotating sphere of gas
into the inner halo and evolve the combined system until $z=0$.
In our models, the gas component is initially hot ($T \sim 10^6 \rm{K}$), with an internal energy structure
determined by the assumption of approximate hydrostatic equilibrium (cf. \citealp{kaufmann, diplom}).
The pressure profile is thus
\begin{equation}
\label{HSE}
p(R)=\int_R^\infty \rho_{gas}\;\frac{G\;M_{\rm{tot}}(r)}{r^2}\;dr,
\end{equation}
where $M_{\rm{tot}}(r)$ is the spherically averaged mass of dark matter and gas within spherical radius $r$.
The gas sphere has a density profile $\rho(r)\propto r^{-1}$ (e.g. \citealp{navarro}) and its total mass is $M_{\rm tot}=5.0-7.5\times10^{10}M_{\odot}$,
in the range of estimated stellar masses for the Milky Way \citep{mcmillan} and of galaxy masses expected for this halo mass range
by abundance matching techniques (e.g. \citealp{guo}). We choose this profile as tests of various other profiles yielded
undesired artifacts. An NFW-like profile, as used by \citet{diplom}, leads to more dominant dispersion-dominated populations forming
from the initially greater central gas mass, which adjusts to the triaxial potential of the central halo before transition
to a disk-like configuration. For a constant density sphere, however, the central build-up of baryonic mass is initially too slow,
delaying the transition from triaxial to disk-like potential relative to the baryonic mass build-up.

The gas sphere is truncated at radius $R_{\rm gas}=60-100\rm \; kpc$. We assume zero pressure outside the gas sphere when determining
the pressure profile according to equation \ref{HSE}. This leads to an initial expansion phase for the outermost gas layers due to the
finite temperatures assigned to the SPH particles. However, this effect is negligible compared to the adjustment of the initially 
spherical gas distribution to the triaxial and substructured gravitational potential of the dark halo, which prevents the setup of
idealized equilibrium initial conditions. The effects of this adjustment are discussed in Sections 3 and 4. Still, as we also show in these
sections, this simple setup is justified by the fact that it leads to disk galaxy models with realistic properties.

We use a baryonic particle mass of $m_{\rm bar}\approx10^5 M_{\odot}$ and a gravitational softening $\epsilon_{\rm{bar}}=205\; \rm{pc}$.
\citet{kaufmann2} found that the angular momentum content and morphology of forming disks depend on resolution. Our resolution
is similar to their highest resolution, for which they found significantly reduced angular momentum losses.
In section 4, we show that angular momentum losses due to lack of resolution are negligible in our simulations.
The gas is set to rotate with a rotation velocity profile
$V_{\rm rot}(r)$, which is either constant or changing linearly from central to outer spherical radii $r$. These profiles yield realistic
disk mass profiles as is shown in the following sections. The angular momentum vector of the gas is aligned either with one of the principal 
axes of the halo or with its angular momentum in order to study how this affects the final 
orientation of the disk (see section 5).

\begin{figure}
\centering
\includegraphics[width=8.5cm]{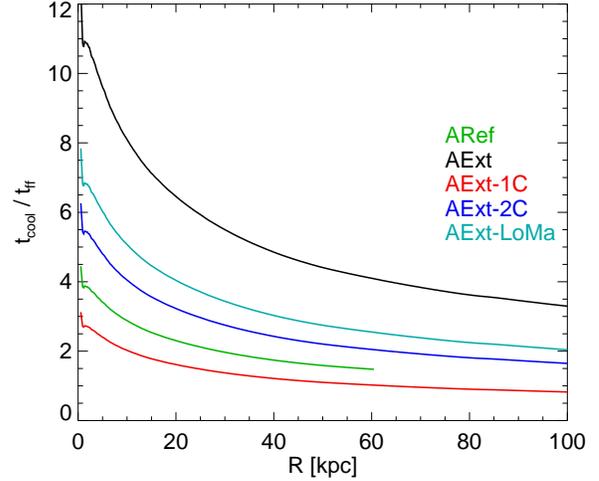}
\caption{The ratio of the cooling time to the free-fall time $t_{\rm{cool}}/t_{\rm{ff}}$ as a function of radius for several models
         with different initial density and cooling time parameters.
         AExt1C, AExt2C and AExt only differ by normalization of the cooling time profile $t_{\rm{cool},0}$.
         AExtLoMa and AExt2C only differ in total gas mass $M_{\rm{gas}}$.
         AExtLoMa and ARef only differ in initial radius of the gas sphere $R_{\rm{gas}}$.
         For a full list of models and their properties see table \ref{overview}.}
\label{tff}
\end{figure}

To model the cooling of the hot gas into the centre of the halo, we apply a simple parametrized cooling function with 
\begin{equation}
t_{\rm cool}=t_{\rm cool, 0} \left( {\rho}\over{\rho_{\rm cool,0}}\right)^{-\alpha},
\end{equation} 
where the normalization $t_{\rm cool, 0}$ determines the rapidity of cooling. We choose $t_{\rm cool,0}=0.5-2.0 \;\rm Gyr$
for $\rho_{\rm cool,0}=10^4 M_{\odot}\; \rm kpc^{-3}\approx4\times 10^{-4} m_{H}\;cm^{-3}$ and $\alpha=0.56$. These values were chosen
to combine a simple dependence of the cooling time on the density with cooling time values which initially are a few times the free fall
time $t_{\rm{ff}}$ at all radii, $t_{\rm{cool}}$ increasing with radius, yet the ratio $t_{\rm{cool}}/t_{\rm{ff}}$ decreasing with radius. 
This enables an initially slow transition of the triaxial dark-matter dominated centre of the halo to a disk-dominated 
quasi-axisymmetric system. Ratios of $t_{\rm{cool}}/t_{\rm{ff}}$ as a function of radius for several models with different
initial density and cooling time parameters are shown in figure \ref{tff} to illustrate how these parameters affect the
cooling of gas in our models. The initial cooling time profiles are affected by the implications of the non-equilibrium initial
conditions as discussed above on dynamical timescales and by the cooling of gas and star formation on cooling timescales. 
In Sections 3 and 4, we show that this cooling time setup produces gas accretion histories appropriate for our models.

A temperature floor $T_{\rm{floor}}=10^5\;\rm{K}$ \citep{kaufmann} is applied to ensure that the forming disk does not 
fragment into a few massive clumps \citep{diplom}, which would then enhance bulge formation. The high choice of 
$T_{\rm{floor}}=10^5\;\rm{K}$ also prevents dense clumps from forming in the infalling gas via the cooling instability \citep{kaufmann}.

\begin{table*}
  \centering
    \begin{tabular}{@{}llrrrrlrlr@{}}
      \hline
      Halo & Model Nr. & $R_{\rm{gas}}$ & $M_{\rm{gas}}$ & $V_{\rm{rot}}$ & $N_{\rm{part}}$ & $t_{\rm{cool}}$ & orientation \\
      \hline
      \hline
      A & ARef       & 60 kpc & $5.0 \times 10^{10} M_{\odot}$ & $120-180 \rm{kms}^{-1}$ & 500000 & 2 & AM300\\
      A & ARef180    & 60 kpc & $5.0 \times 10^{10} M_{\odot}$ & $120-180 \rm{kms}^{-1}$ & 500000 & 2 & -AM300\\
      A & ARef-Min   & 60 kpc & $5.0 \times 10^{10} M_{\odot}$ & $120-180 \rm{kms}^{-1}$ & 500000 & 2 & -Minor\\
      A & ARef+Min   & 60 kpc & $5.0 \times 10^{10} M_{\odot}$ & $120-180 \rm{kms}^{-1}$ & 500000 & 2 & +Minor \\
      A & ARef-Med   & 60 kpc & $5.0 \times 10^{10} M_{\odot}$ & $120-180 \rm{kms}^{-1}$ & 500000 & 2 & -Medium \\
      A & ARef+Med   & 60 kpc & $5.0 \times 10^{10} M_{\odot}$ & $120-180 \rm{kms}^{-1}$ & 500000 & 2 & +Medium \\
      A & ARefEnd    & 60 kpc & $5.0 \times 10^{10} M_{\odot}$ & $120-180 \rm{kms}^{-1}$ & 500000 & 2 & EndARef\\
      A & ARef+Maj   & 60 kpc & $5.0 \times 10^{10} M_{\odot}$ & $120-180 \rm{kms}^{-1}$ & 500000 & 2 & +Major\\
      A & ARef-Maj   & 60 kpc & $5.0 \times 10^{10} M_{\odot}$ & $120-180 \rm{kms}^{-1}$ & 500000 & 2 & -Major \\
      A & ARef45     & 60 kpc & $5.0 \times 10^{10} M_{\odot}$ & $120-180 \rm{kms}^{-1}$ & 500000 & 2 & 45\\
      \hline
      A & AExt       & 100 kpc & $7.5 \times 10^{10} M_{\odot}$ & $120 \rm{kms}^{-1}$ & 750000 & 4 & AM300 \\
      A & AExt1C     & 100 kpc & $7.5 \times 10^{10} M_{\odot}$ & $120 \rm{kms}^{-1}$ & 500000 & 1 & AM300\\
      A & AExt2C     & 100 kpc & $7.5 \times 10^{10} M_{\odot}$ & $120 \rm{kms}^{-1}$ & 750000 & 2 & AM300\\
      A & AExtMedAM  & 100 kpc & $7.5 \times 10^{10} M_{\odot}$ & $80 \rm{kms}^{-1}$ & 750000 & 4 & AM300 \\
      A & AExtLoAM   & 100 kpc & $7.5 \times 10^{10} M_{\odot}$ & $40 \rm{kms}^{-1}$ & 750000 & 4 & AM300\\
      A & AExt-Med   & 100 kpc & $7.5 \times 10^{10} M_{\odot}$ & $120 \rm{kms}^{-1}$ & 500000 & 4 & -Medium \\
      A & AExt180    & 100 kpc & $7.5 \times 10^{10} M_{\odot}$ & $120 \rm{kms}^{-1}$ & 750000 & 4 & -AM300\\
      A & AExtLoMa   & 100 kpc & $5.0 \times 10^{10} M_{\odot}$ & $120 \rm{kms}^{-1}$ & 500000 & 2 & AM300 \\   
      A & AExtHiMa   & 100 kpc & $15.0 \times 10^{10} M_{\odot}$ & $120 \rm{kms}^{-1}$ & 750000 & 4 & AM300\\
     \hline
      A & AMaj       & 60 kpc & $5.0 \times 10^{10} M_{\odot}$ & $60 \rm{kms}^{-1}$ & 500000 & 4 & +Major\\
      A & AMaj2C     & 60 kpc & $5.0 \times 10^{10} M_{\odot}$ & $60 \rm{kms}^{-1}$ & 500000 & 2 & +Major \\
      A & AMaj180    & 60 kpc & $5.0 \times 10^{10} M_{\odot}$ & $60 \rm{kms}^{-1}$ & 500000 & 4 & -Major\\
      \hline
      A & ACosmo     & 70 kpc & $5.0 \times 10^{10} M_{\odot}$ & $120 \rm{kms}^{-1}$ & 500000 & 2 & Cosmo\\
      \hline
      \hline
      C & CExt+AM    & 100 kpc & $7.5 \times 10^{10} M_{\odot}$ & $120 \rm{kms}^{-1}$ & 750000 & 4 & AM300\\
      C & CExt-AM   & 100 kpc & $7.5 \times 10^{10} M_{\odot}$ & $120 \rm{kms}^{-1}$ & 750000 & 4 & -AM300\\
      C & CExt+Min  & 100 kpc & $7.5 \times 10^{10} M_{\odot}$ & $120 \rm{kms}^{-1}$ & 750000 & 4 & +Minor\\
      C & CExt-Min  & 100 kpc & $7.5 \times 10^{10} M_{\odot}$ & $120 \rm{kms}^{-1}$ & 750000 & 4 & -Minor \\
      C & CExt+Med  & 100 kpc & $7.5 \times 10^{10} M_{\odot}$ & $120 \rm{kms}^{-1}$ & 750000 & 4 & +Medium\\
      C & CExt-Med  & 100 kpc & $7.5 \times 10^{10} M_{\odot}$ & $120 \rm{kms}^{-1}$ & 750000 & 4 & -Medium \\
      C & CExt-Arb  & 100 kpc & $7.5 \times 10^{10} M_{\odot}$ & $120 \rm{kms}^{-1}$ & 750000 & 4 & -CArb\\
      C & CExt+Arb  & 100 kpc & $7.5 \times 10^{10} M_{\odot}$ & $120 \rm{kms}^{-1}$ & 750000 & 2 & CArb\\
      C & CExtEnd   & 100 kpc & $7.5 \times 10^{10} M_{\odot}$ & $120 \rm{kms}^{-1}$ & 750000 & 4 & EndCExt+Min\\
      C & CExtCosmo & 100 kpc & $7.5 \times 10^{10} M_{\odot}$ & $120 \rm{kms}^{-1}$ & 750000 & 4 &  Cosmo\\
      \hline
      C & C+Maj      & 60 kpc & $5.0 \times 10^{10} M_{\odot}$ & $60 \rm{kms}^{-1}$ & 500000 & 4 & +Major\\
      C & C-Maj      & 100 kpc & $5.0 \times 10^{10} M_{\odot}$ & $60 \rm{kms}^{-1}$ & 500000 & 4 & -Major \\
      \hline
      \hline
    \end{tabular}
    \caption{\textbf{Overview over our models}
            \newline
             \textit{Column 1}:halo A or C;
             \textit{Column 2}:our model name;
             \textit{Column 3}:the radius of the initial gas sphere;
             \textit{Column 4}:the mass of the initial gas sphere;
             \textit{Column 5}:the rotational velocity of the initial gas sphere, a single value stands for a constant $V_{\rm{rot}}(r)$,
                               $a-b \;\rm{kms}^{-1}$ stands for a linear increase in $V_{\rm{rot}}(r)$ from $a$ in the 
                               centre to $b$ at $R_{\rm{gas}}$;
             \textit{Column 6}:the number of baryonic particles;              
             \textit{Column 7}:the cooling time profile applied; the number stands for the approximate ratio $t_{\rm{cool}}/t_{\rm{dyn}}(R_{\rm{gas}})$
                               (see text and figure \ref{tff});
             \textit{Column 8}:the orientation of the model. AM300 stands for the orientation of the halo angular momentum within 300 kpc,
                               Major, Medium and Minor stand for the principal axes of the halo, `+' and `-' are arbitrarily chosen and differ
                               by 180 degrees, EndARef stands for the $z=0$ orientation of model ARef, 45 stands for an orientation halfing the angle
                               between Major and Minor, Cosmo stands for the orientation of galaxy angular momentum in the cosmological runs of
                               CS09, CArb is an arbitrary orientation in halo C, EndCExt+Min stands for the $z=0$ orientation of model 
                               CExt+Min.}
  \label{overview}
\end{table*}

Such a setup clearly ignores evidence from cosmological hydrodynamical simulations that steady, narrow, cold gas streams
penetrating the shock-heated atmosphere of massive dark matter haloes, are important in feeding gas
onto forming galaxies (e.g. \citealp{dekel}). However, cold streams are only expected to be the dominant accretion mode at $z>2$
(\citealp{faucher}) and even at these redshifts, there is little direct observational evidence for them (e.g. \citealp{steidel}).
The majority of the stellar mass in the Milky Way disk was formed after $z\sim 2$ \citep{ab} and
it may be argued that the underlying continuous star formation since $z\sim1$ is driven by cooling of hot coronal gas in the wakes of 
galactic fountain clouds, which transfers gas from the virial-temperature corona to the disk (e.g. \citealp{mari}). 
Our focus here is not on simulating a realistic assembly process, but rather on using a simplified scheme,
which allows important aspects of the underlying dynamics to be clarified.

In our simulations, once the gas has reached the temperature floor and crossed a density threshold $\rho_{0,\rm sfr}$, it can form stars with
\begin{equation}
t_{\rm sfr}=t_{\rm sfr,0}\left( {\rho}\over {\rho_{0,\rm sfr}}\right)^{-\beta}.
\end{equation} 
We choose $\rho_{0, \rm{sfr}}$ to correspond to a hydrogen number density $n_{\rm{H, thresh}}=0.1\;\rm{ cm^{-3}}$ and
we use $\beta=0.5$ and $t_{\rm{sfr,0}}= 10 \times t_{\rm{dyn}}(\rho_{0,\rm sfr})$, as often applied
in simulations of this resolution (CS09). Gas particles eligible for star formation are stochastically
turned into collisionless particles of the same mass (see e.g. \citealp{lia}). 
In our simulation, the formation of stars is not accompanied by any kind of feedback processes.
We have run tests with 50 times higher values of $\rho_{0,\rm sfr}$ and/or 10 times lower values of $T_{\rm{floor}}$ 
and did not find any major changes to our results. Both changes tend to produce dynamically hotter stellar disks
as a lower cooling floor leads to stronger fragmentation of the disk and thus enhances disk heating by dense substructures.
A higher density threshold lowers star formation rates in the later, gas-poor evolution of our disk models thus diminishing
the young, cold disk component.

\section{Quantifying a disk model}

\begin{figure*}
\centering
\includegraphics[width=17cm]{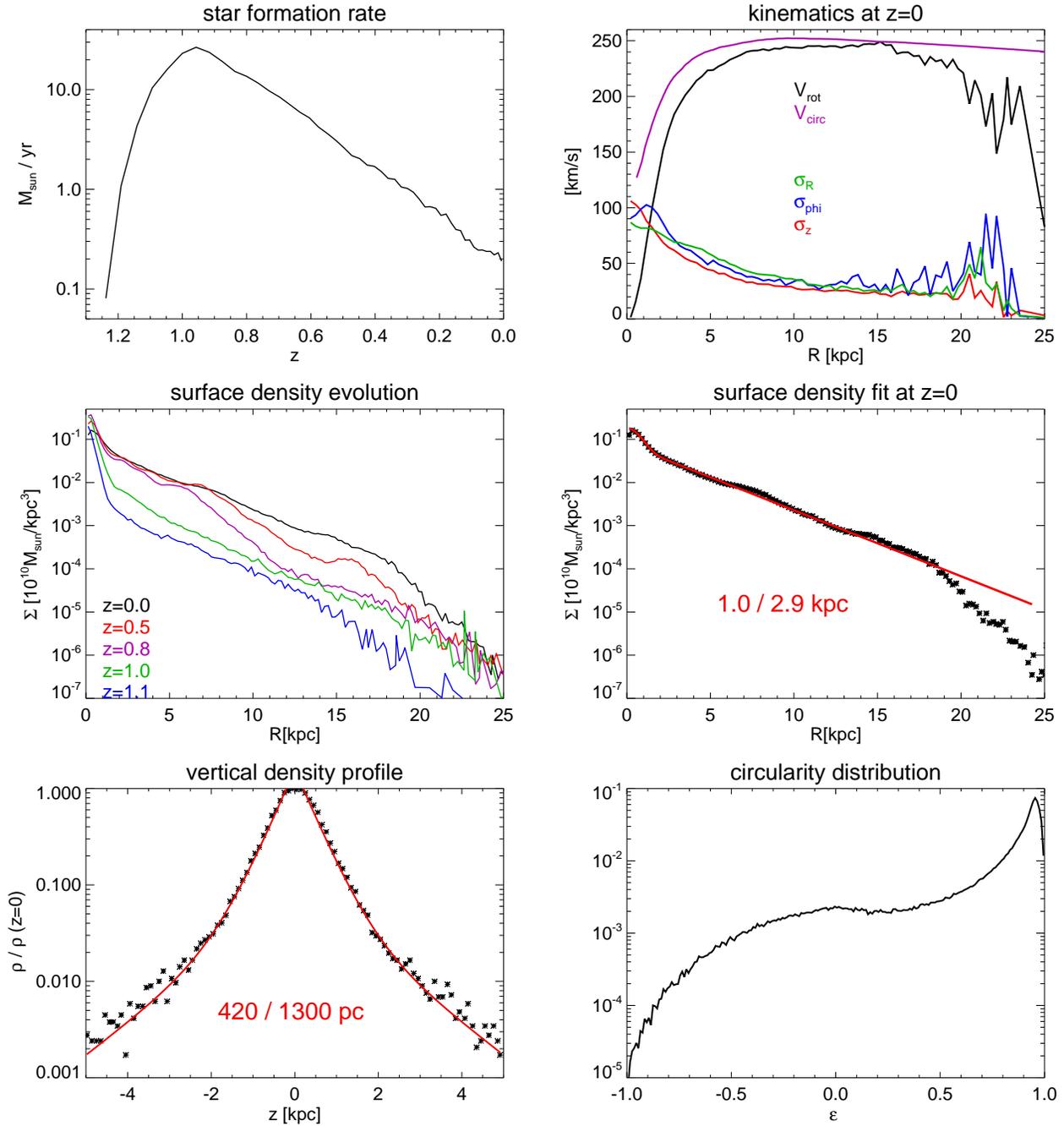}
\caption{Properties of model ARef.
         \textit{Top Left}: Star formation rate as a function of redshift $z$. 
         \textit{Top Right}: Stellar kinematics as a function of disk radius $R$ at $z=0$. 
                           Only stars within 500 pc of the midplane were taken into account.
                           In purple we overplot the spherically-averaged circular velocity
                           $V_{\rm{circ}}=\sqrt{{GM(<r)}\over{r}}$.
         \textit{Middle Left}: The evolution of the surface density profile $\Sigma(R)$ from $z=1.1$ to $z=0$.
         \textit{Middle Right}: Surface density profile $\Sigma(R)$ at z=0 fitted with a two-component profile
                           $ \Sigma_{0,1} \; \exp \left( - {{R}\over{R_d}} \right)
                           + \Sigma_{0,2} \; \exp \left[ - \left( {R} \over {R_b} \right)
                           ^{\left({1} \over {n}\right)} \right] $ with scalelengths $R_d = 2.9 \;\rm{kpc}$
                           and $R_b = 1.0 \;\rm{kpc}$ and $n=0.47$.
         \textit{Bottom Left}: Normalized $z=0$ vertical density profile at $R=5\;\rm{kpc}$ fitted by a (double-)exponential with
                           $h_{\rm{thin}}=420\;\rm{pc}$ and  $h_{\rm{thick}}=1300\;\rm{pc}$                        
         \textit{Bottom Right}: Circularity distribution for all stars at $z=0$.                                   
         }
\label{ARef}
\end{figure*}

We start the analysis of our models by discussing in detail one model in halo A, which is named ARef.
We choose this model as it produces a system which has certain structural similarities to the Milky Way.
The parameters applied are an initial radius of $R_{\rm{gas}}=60\;\rm{kpc}$ , a total mass of $M_{\rm{gas}}=5\times 10^{10}M_{\odot}$,
a cooling time which is about twice as long as the free-fall time for the outermost gas (see figure \ref{tff}), a rotational velocity
which increases from $120\;\rm{kms^{-1}}$ in the centre to $180\;\rm{kms^{-1}}$ at $R_{\rm{gas}}$ and an orientation of the baryonic
angular momentum vector parallel to that of the dark matter within $300\;\rm{kpc}$.
This orientation is close to the minor axis of the potential of the inner halo.
The dependence of the final disk properties properties on these initial parameters is discussed in the following section.
We split this section into subsections that deal with disk structure, angular momentum loss, kinematics
and disk heating.

\subsection{Structural evolution}

Analyzing ARef at $z=0$, $86 \%$ of the baryonic mass has formed stars and the remaining gas is situated mainly in a thin, extended gas disk (see 
top row in figure \ref{sd2d}). This gas fraction is in reasonable agreement with $z \sim 0$ spirals \citep{evoli}. As can be seen in the top left
panel of figure \ref{ARef}, the star formation rate (SFR) rises steeply to a peak at $z\sim1$ before falling by two orders of magnitude
by $z=0$. 

In the middle left panel of figure \ref{ARef} we show the evolution of the stellar surface density profile $\Sigma(R)$ of model ARef from $z=1.1$ until
$z=0$. It can be seen that the first stars form in a concentrated bulge-like structure at $R<2\;\rm{kpc}$. This structure is actually triaxial
because the infalling gas adjusts to the potential imposed by the halo. Its surface density only increases by a factor of
$\sim 2$ by $z=0$. The surface density profile outside the bulge region is already exponential in this early phase.
Once most of the baryons have cooled, the infalling gas and the forming stars settle in a well-defined disk, which 
grows transforming the central potential into an axisymmetric configuration.
The disk grows inside-out showing a truncation at a radius $R_{\rm{max}}(z)$ as observed in disk galaxies \citep{trunc2}.
This is a consequence of high-angular-momentum gas living initially at greater radii, lower densities
and thus longer cooling times. Moreover, according to our assumed star formation law, star formation timescales are longer
at lower surface densities. The break in $\Sigma(R)$ moves outward with time to around $R\sim 19\;\rm{kpc}$ at $z=0$.
There is observational evidence for inside-out formation of real disk galaxies \citep{wang} 
and for outward-movement of their truncations \citep{trunc}.

Due to the numerically imposed cooling floor $T_{\rm floor}$ and the density threshold $\rho_{\rm{thresh}}$, there is
an imposed star formation threshold surface density $\Sigma_{\rm{thresh}}$. As is shown in the first row of figure \ref{sd2d}
the gas disk at $z=0$ extends to larger radii, where this threshold is not reached. 
This also indicates that the truncation of the initial spherical distribution does not play a role here.
The extended gas disk is also slightly warped, as in real disk galaxies \citep{sancisi}. The warp is a result of a slight misalignment
between the disk and the late-infall gas (see sections 5 and 7 for a discussion of misalignments).

Exponential disk profiles with central concentrations are common in real galaxies \citep{devau} and also in simulations \citep{katz},
although most simulations significantly over-predict the mass in the central bulge component (CS09, \citealp{cs2010},
but see \citealp{4disk} for simulations of lower mass (nearly) bulge-less disks).
The $\Sigma(R)$ profile at $z=0$ as displayed in the middle right panel of figure \ref{ARef} is well-fitted by a profile of the form
$ \Sigma_{0,1} \; \exp \left( - {{R}\over{R_d}} \right) + \Sigma_{0,2} \; \exp \left[ - \left( {R} \over {R_b} \right)
 ^{\left({1} \over {n}\right)} \right] $, which combines an exponential disk with a Sersic profile. The Sersic-indices
found for our disk models are $n\sim 0.5-2.$, similar to observed pseudo-bulges. The scale-lengths for ARef
are $R_b=1.0\;\rm{kpc}$ and $R_d= 2.9 \; \rm{kpc}$. Calculating a disk-to-bulge mass ratio from this profile alone yields $D/B\sim 12$.

In the bottom left panel of figure \ref{ARef} we display the vertical stellar density profile $\rho(z)$ for all stars in model ARef that are located
in an annulus of width 1 kpc centred around $R=5\;\rm {kpc}$ at $z=0$. The profile is well fitted
by a double exponential with scale heights 420 and 1300 pc. The vertical profile does not vary significantly within $R<10\;\rm{kpc}$,
but at larger radii the disk flares and the scale height increases by a factor of $\sim 2$ until $R_{\rm{max}}$.
Thus the outer vertical structure of this model is in conflict with observations of constant vertical structure
within the exponential disk region \citep{trunc2}. In the following section we will show that this conflict is absent for some of our models.
The flaring and the `thick' disk are caused by old stellar material formed in the bulge-formation period around $z\sim 1$.
From then on the scale-heights first become smaller due to the growth of a young, cold disk before beginning to increase again due
to disk heating. This increase is stronger in the inner disk as the outer disk has a younger mass-weighted age.
The ARef disk thus has similar structural characteristics to the Milky Way disk \citep{juric, mcmillan}, although its disk is $\sim40$ \% thicker.

\subsection{Angular momentum loss of infalling gas}

\begin{figure}
\centering
\includegraphics[width=8.5cm]{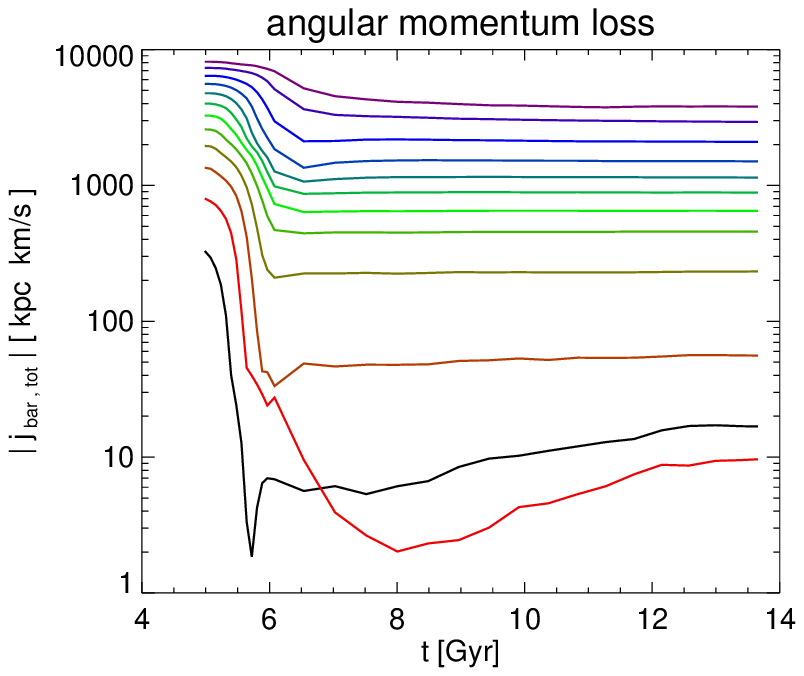}
\caption{Angular momentum loss as a function of time for model ARef. The initial gas sphere has been divided into 12 spherical shells
         of 5 kpc width each. For each shell the absolute of the total angular momentum vector $\bf{j}_i$ of all particles initially
         in shell \textit{i} is calculated at each output.}
\label{aml}
\end{figure}

The radial structure of the disk discussed above is determined by the potential and the angular momentum content of the stellar population.
The concentrated peak in $\Sigma(r)$ produced in the early stages of star formation indicates an unexpectedly high amount of low angular momentum
material and thus angular momentum losses. We investigate this in figure \ref{aml}, in which we plot the evolution of the total specific
angular momentum of populations of baryons initially located in concentric shells of 5 kpc width.
The material initially located at $R<10\;\rm{kpc}$ loses $>\sim 99 \%$ of its angular momentum within the first 500 Myr.
The material in the next two shells still loses $>\sim 90 \%$, the outermost material $\sim 50 \%$.
The strongest losses occur in the phase when the central potential is transformed from a triaxial to a disk-like configuration, in the
process of which the bulge forms.
Moreover, the losses are subsequently delayed for outer shells, indicating that the angular momentum
is lost as the gas cools and moves to the centre.

\subsection{Kinematical properties}

We move on to analyze the kinematics of ARef.
The top right panel of figure \ref{ARef} depicts the rotation velocity $V_{\rm {rot}}$ and the velocity dispersions
$\sigma_R$, $\sigma_{\phi}$ and $\sigma_z$ of the disk stars and the spherically averaged circular velocity
$V_{\rm{circ}}(r)=\sqrt{{GM(r)}/{r}}$ profile at $z=0$. $V_{\rm{circ}}$ peaks at $\sim 250 \;\rm{kms^{-1}}$ at $R\sim 10\;\rm{kpc}$
and gently decreases outside. The stellar rotation curve shows a broad peak at $7 < R/\rm{kpc} < 15$ at $V_{\rm {rot}}\sim 240\; \rm {kms^{-1}}$, 
For velocity dispersions, we find a decreasing profile out to $R\sim 12\;\rm{kpc}$ and $\sigma_R > \sigma_{\phi} >\sigma_z$ in the
rotation-dominated region. The kinematical properties of the ARef disk are thus also similar to , but slightly hotter than the Milky Way
disk \citep{ab, ralph2012}.

As a tool to quantify disk-to-bulge ratios, the circularity $\epsilon=V_{\rm{rot}}/V_{\rm {rot,max}}(E)$ for a stellar particle
at energy $E$ has been widely used \citep{abadi}. Particles on a perfect circular orbit have $\epsilon=1$ and thus a disk will show
a peak close to $\epsilon=1$, whereas a non-rotating bulge will have a peak at $\epsilon=0$. The bottom right panel of 
figure \ref{ARef} depicts the $\epsilon$-histogram
for ARef, which shows a distinct peak at $\epsilon=0.95$ and only a shallow peak at $\epsilon=0$, corresponding to a $D/B\sim 4$. Thus
the kinematic estimate for $D/B$ is significantly lower than the one from $\Sigma (R)$-fitting, consistent with e.g. the results of
\citet{cs2010} for simulations of disk formation from cosmological initial conditions.

\begin{figure*}
\centering
\includegraphics[width=8.5cm]{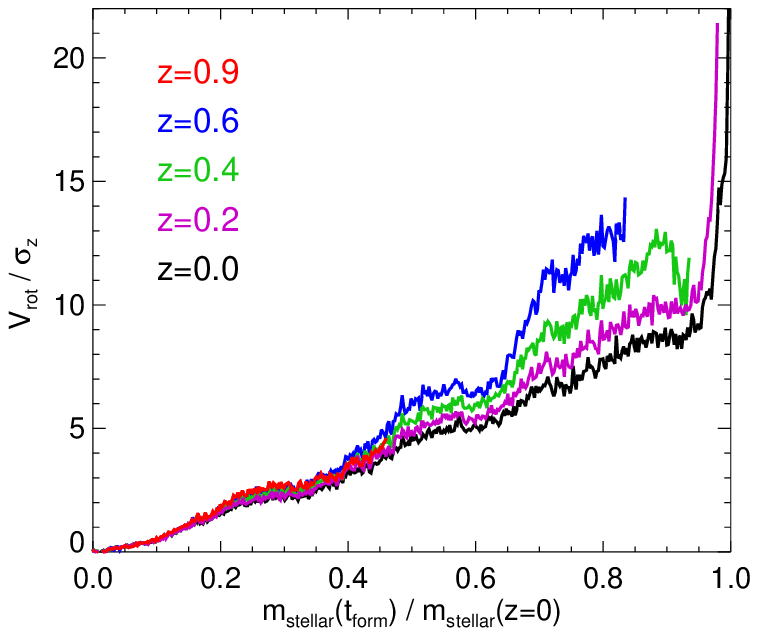}
\includegraphics[width=8.5cm]{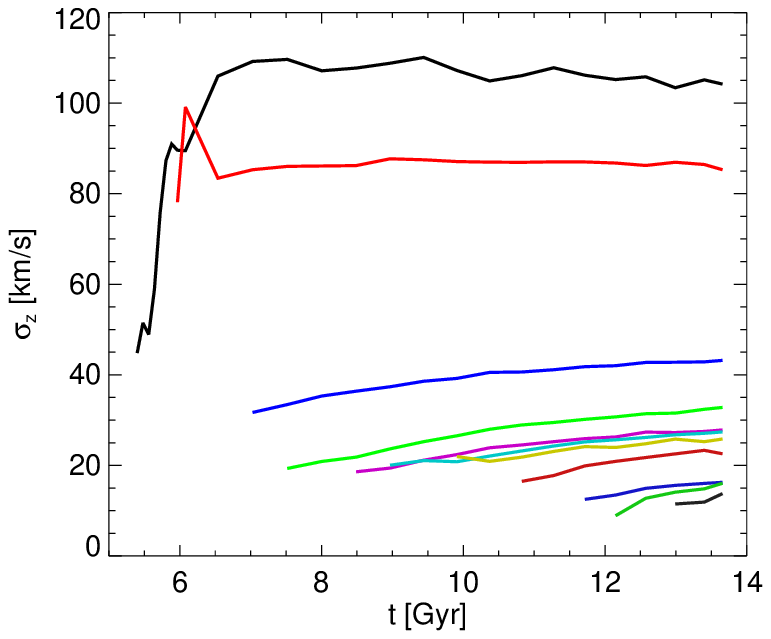}
\caption{Disk Heating in model ARef: 
         \textit{Left}: The ratio of the rotation velocity of stars to their vertical velocity dispersion $V_{\rm{rot}}/\sigma_z$ as a function
         of the fraction of stellar mass that has formed until the formation time $t_{\rm{form}}$ of a star particle relative to the
         total stellar mass at z=0 for model ARef. The first stars are found at x-value 0, the ones formed at
         $z=0$ at x-value 1. Curves are plotted for redshifts $z=$ 0.9, 0.6, 0.4, 0.2 and 0. The endpoints of curves indicate the
         fraction of stellar mass that has formed until the given redshift, e.g. 82 \% for $z=0.6$.
         The evolution of $V_{\rm{rot}}/\sigma_z$ over time can be inferred by considering the various colored lines at one specific x-value.
         At each output redshift $z$ we define a disk plane perpendicular to the total stellar angular momentum.
         $V_{\rm{rot}}$ is then calculated as the mean tangential velocity and $\sigma_z$ as the rms velocity in 
         the direction of the rotation axis of all stars in a certain age bin. 
         \textit{Right}: Vertical velocity dispersion $\sigma_z(t)$ as a function of time for coeval populations in model ARef.
         The stellar population is binned into 50 equally spaced bins in redshift $z$, a representative selection of which is
         depicted by the various colors in the figure. The starting point of a line indicates the formation time of the population
         in consideration.
}
\label{vs1}
\end{figure*}

However, this analysis is not capable of distinguishing between old and young stars and is also not good at quantifying how thin
and dynamically cold a disk is. In the left panel of figure \ref{vs1} 
we therefore employ a plot of the rotation-to-dispersion ratio $V_{\rm rot}/\sigma_z$
as a function of the age of the stars. 
We define a disk plane perpendicular to the total stellar angular momentum at each output time $t$.
$V_{\rm{rot}}$ is calculated as the mean tangential velocity and $\sigma_z$ as the rms velocity in the direction of the rotation axis
of all stars in a certain age bin. 
Keep in mind that the Milky Way young thin disk has $V_{\rm rot}/\sigma_z > \sim 20$, the older 
thin disk has $\sim 10$ \citep{ab}, whereas the thick disk shows $\sim 5$ (\citealp{reddy}).
To be able to compare different models better, we sort the stars according to their
age, bin them in equal mass bins, and sort those according to the fraction of the $z=0$ stellar mass that has formed up to the time in consideration.
We do this for five different redshifts. The black $z=0$ line shows that $V_{\rm rot}/\sigma_z$ monotonically decreases with increasing age.
Only the youngest stars are compatible with thin disk kinematics. The oldest $\sim 1/5$ is dispersion dominated and corresponds to the
bulge peak seen in the bottom left panel of figure \ref{ARef}. Most of the galaxy thus has thick disk kinematics.
Considering the evolution of $V_{\rm rot}/\sigma_z$ over time, we can see that the bulge kinematics do not change from $z=0.9$ to $z=0.0$, whereas
the disk populations show decreasing $V_{\rm rot}/\sigma_z$ ratios over time, i.e. the disk populations are heated, indicating that
at least $1/3$ of the stellar particles are born with thin-disk-like kinematics.

\subsection{Disk heating}

In order to analyze disk heating as it occurs in ARef in more detail, we plot the evolution
of $\sigma_z$ for several coeval populations through time in the right panel of figure \ref{vs1}.
The stars that form earliest and most centrally (see middle left panel in figure \ref{ARef})
show high, almost isotropic velocity dispersion (top right panel in figure \ref{ARef}). 
These stars are strongly heated within a few $10^8$ years (black line) or born hot (red line).
Within $\sim 1 \;\rm{Gyr}$ after the first star formation in the models the initial $\sigma_z$ has already become significantly lower,
indicating the transition from bulge to disk formation. Within another $\sim 0.5 \;\rm{Gyr}$ $\sigma_z$ at birth has dropped below
$20 \;\rm {kms^{-1}}$. Only the stars that form latest form with a smaller $\sigma_z \sim 10 \;\rm {kms^{-1}}$.
As will be shown below, the high initial velocity dispersions of the first stars in this model 
are an effect of the structure of the dark halo. Test simulations
in spherically symmetric haloes do not show this effect, but a significantly reduced initial velocity dispersion of the stars that form
first, of the order $\sim 10-20 \rm {kms}^{-1}$, as seen for the young populations in model ARef.
The velocity dispersion of stellar particles in numerical simulations is limited by the applied star formation threshold, resolution and 
modeling of gas physics \citep{house} and the formation of a realistically cold
young thin disk population is thus not possible in our current models.

All the disky coeval populations show an increase of $\sigma_z$ over time. Such continuous disk heating has been inferred for
solar neighbourhood stars \citep{gcs} and is usually linked to scattering of stars off disk substructure
such as transient spirals and/or giant molecular clouds \citep{jenkins}. For our models we cannot, however, exclude that this
effect is partly due to numerical heating (e.g. \citealp{steinmetz}). A closer analysis of the phenomenon over all of our models
reveals that the strongest heating always occurs when distinct substructure in the form of a bar or spirals is present.
A look at the evolution of $\sigma_z$ at different disk radii $R$ over time reveals that due to the combined effect of the heating
of coeval populations and the inside-out formation of our models in the inner disk regions, 
$\sigma_z$ first drops due to the formation of a thin disk and then increases due to disk heating, whereas at outer radii the
formation phase lasts longer and $\sigma_z$ remains almost constant after an initial drop.

\subsection{Summary}

In conclusion, our setup is capable of producing disk galaxy models with realistic structural and kinematic properties.
Disk formation proceeds inside-out and follows an early bulge-formation episode, during which the central halo potential
is transformed into a disk-like configuration. This transformation is connected to angular momentum losses for the infalling gas.
Moreover, a thick disk component in the vertical profiles originates from this epoch.
We confirm that for a disk/bulge decomposition, a kinematic analysis is superior to a surface-density decomposition.
However, circularity distributions alone, as often applied, are not ideally suited to distinguish properly between thin and thick disk components,
which is why $V_{\rm{rot}}/\sigma_z$ ratios should be used in addition. Our model does show a significant thin disk component, however its 
$V_{\rm{rot}}/\sigma_z$ is limited to values $\sim 10$ compared to 20 for real young thin disks,
due to the numerical lower limit we impose on $\sigma_z$.
The disk undergoes secular disk heating and shows a significant decline of zero-age velocity dispersion of stars.

\begin{figure}
\centering
\includegraphics[width=8.5cm]{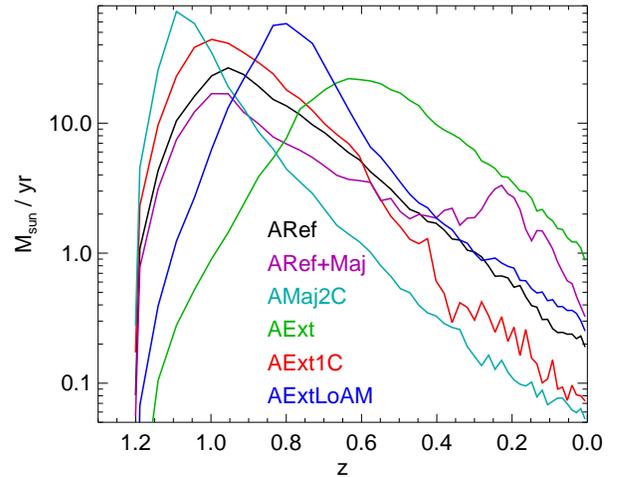}
\caption{Star formation rates as a function of redshift $z$ for various model.}
\label{sfr}
\end{figure}

\begin{figure*}
\centering
\includegraphics[width=8.5cm]{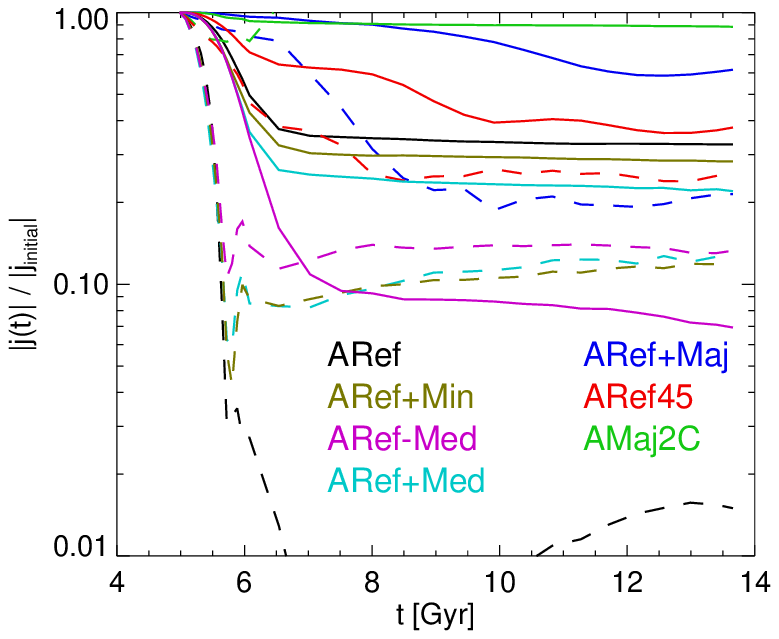}
\includegraphics[width=8.5cm]{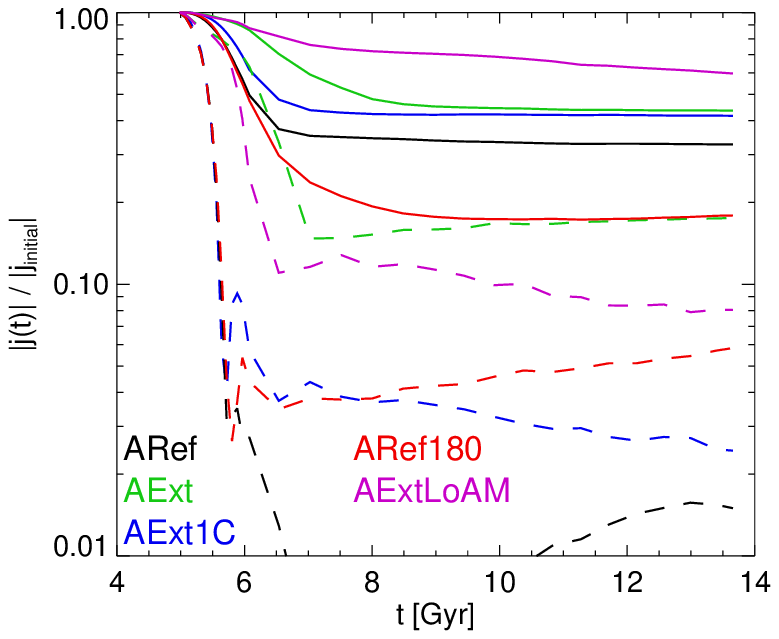}
\caption{Angular momentum loss for different models. The left panel features different orientations with otherwise constant parameters, whereas
the right panel focuses on different representations of physics at the same orientation. 
The definition of angular momentum loss is as in figure \ref{aml}.
The solid lines represent all baryonic particles in the simulations, the dashed lines only the particles that initially reside at $R<0.2\;R_{\rm{gas}}$.
}
\label{angstuff}
\end{figure*}

\section{How to make better disks}

In this and the next section we discuss the dependencies of our models on the parameters that define them:
\begin{enumerate}
  \item The mass of the model $M_{\rm{gas}}$, the initial radius $R_{\rm{gas}}$ and thus the initial density profile $\rho(R)$
  \item The rapidity of cooling and thus accretion of gas onto the forming galaxy represented by the cooling time normalization $t_{\rm {cool,0}}$
  \item The initial angular momentum distribution of the gas represented by the rotation velocity profile $V_{\rm{rot}}(R)$
  \item The orientation of the angular momentum vector of the gas (see also the following section)
\end{enumerate}
Our goal is to understand which physical conditions favor thin disk formation in $\Lambda$CDM haloes.
An overview over our models and their parameters can be found in Table \ref{overview}. In this section, we will mainly focus on
halo A. For the parameters $M_{\rm{gas}}$, $R_{\rm{gas}}$, $t_{\rm {cool,0}}$ and $V_{\rm{rot}}(R)$ we could not find significant
halo-to-halo differences. Most differences because of orientations are discussed in Section 5.
As in the previous section, we split our analysis into subsections that deal with star formation,
angular momentum loss, disk structure, kinematics and disk heating.

\subsection{Gas infall and star formation}

We begin with a discussion of how the model parameters influence the formation histories of our models.
The star formation rates of the models displayed in figure \ref{sfr} reveal that the process of gas inflow depends mainly on $t_{\rm {cool,0}}$
and the loss of angular momentum, which is put into the initial models
via $V_{\rm{rot}}(R)$. Models ARef (black line), ARef+Maj (purple) and AMaj2C (turquoise) 
share the initial density profile and cooling time. All models show a steep rise followed by a softer decrease in SFR.
ARef+Maj and ARef differ only in orientation, consequently they show qualitatively similar SFRs.
ARef+Maj loses a smaller amount of angular momentum (see below), thus forms a more extended structure with lower densities and has longer star formation 
timescales, a lower peak SFR and a higher late-time SFR. The second peak for ARef+Maj corresponds to star formation in a ring structure
(see the following section).
ARef+Maj and AMaj2C differ only in $V_{\rm{rot}}(R)$ with AMaj2C having 
significantly less angular momentum, which leads to a much more concentrated object,
higher densities, higher cooling rates, higher SFRs, an earlier peak and a steeper decrease in SFR.
The same is true for a comparison of AExt1C (red line), AExt (green) and AExtLoAM (blue). They are all more extended with a higher
initial gas mass than the models
mentioned so far and thus represent a different initial density profile. AExt1C cools 4 times faster than AExt and AExtLoAM has 3 times less
angular momentum than AExt. In both cases this results in a higher and earlier peak in SFR.

Directly connected to the SFR is the remaining gas fraction in the disks. At $z=0$ these are almost independent
of cooling time, as the cooling and star formation timescales are small compared to the simulation time and the gas disk outside $R_{\rm{max}}$
contains  a negligible amount of mass. Thus AExt1C and AExt both show a final gas fraction of 18\%. Angular momentum variation can significantly
alter the mass in the outer gas disk with very long cooling and SF timescales. Thus, ARef+Maj (36\%) and AExt (18\%) have much higher
gas fractions than AMaj2C (3\%) and AExtLoAM (4\%). Realistic gas ratios are 5 to 25\% \citep{evoli}, so that only our most extreme models 
are inconsistent with observations.

\subsection{Why does infalling gas lose angular momentum?}

As we have shown in Section 3.2, gas in our models loses angular momentum as it falls to the centre.
Figure \ref{angstuff} allows a closer inspection of angular momentum loss in several models. For each model
the solid line represents the absolute value of the total specific angular momentum of all baryonic particles, whereas the dashed line represents only
the particles that initially live within $R<0.2\;R_{\rm{gas}}$. The left panel focuses on models which differ only in orientation and apart from 
AMaj2C, share the same density and velocity profiles and cooling times. Apart from ARef-Med and AMaj2C the central parts show higher angular 
momentum losses. The difference between central and total mass is biggest for model ARef as discussed above. Focusing on total angular
momentum losses, the models aligned with the major halo axis (ARef+Maj and AMaj2C) show the lowest losses. They are about 10\% at $t= 8\; \rm {Gyr}$
and subsequently only increase for ARef+Maj which develops strong misalignments between inner and outer galaxy, leading to higher losses than in 
AMaj2C, which forms a well-aligned disk. ARef45, which starts with its orientation 45 degrees offset, has lost 40\% 
of its total angular momentum at $t= 8\; \rm {Gyr}$. The rest 
of the models, which all started oriented perpendicularly to the minor axis, show losses well in excess of 50\%. 
Only a small part of these losses, perhaps of the order of 10\% can be of numerical origin.
\citet{kaufmann2} reported 10\% losses for their highest resolution model in a spherical halo, the 
resolution of which is similar to our models.

\begin{figure}
\centering
\includegraphics[width=8.5cm]{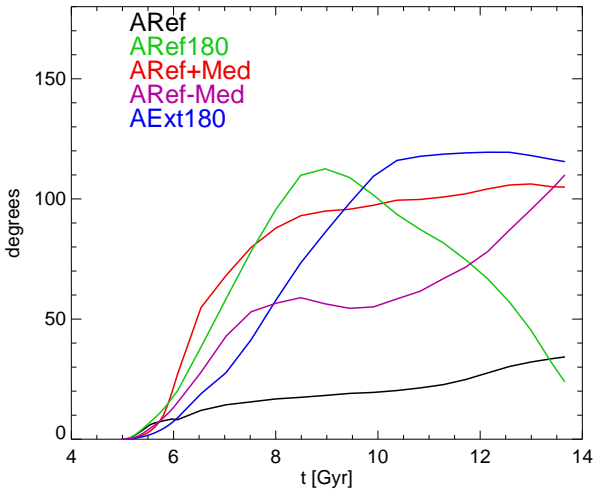}
\caption{Mass-weighted reorientation angle of all baryons as a function of time for models ARef-Med, ARef+Med, ARef, ARef180, AExt180.
         The angle is measured in a fixed coordinate system between the initial angular momentum vector orientation at $z=1.3$ and
         its given orientation at time $t$.
} 
\label{reorient}
\end{figure}

\begin{figure*}
\centering
\includegraphics[width=14. cm]{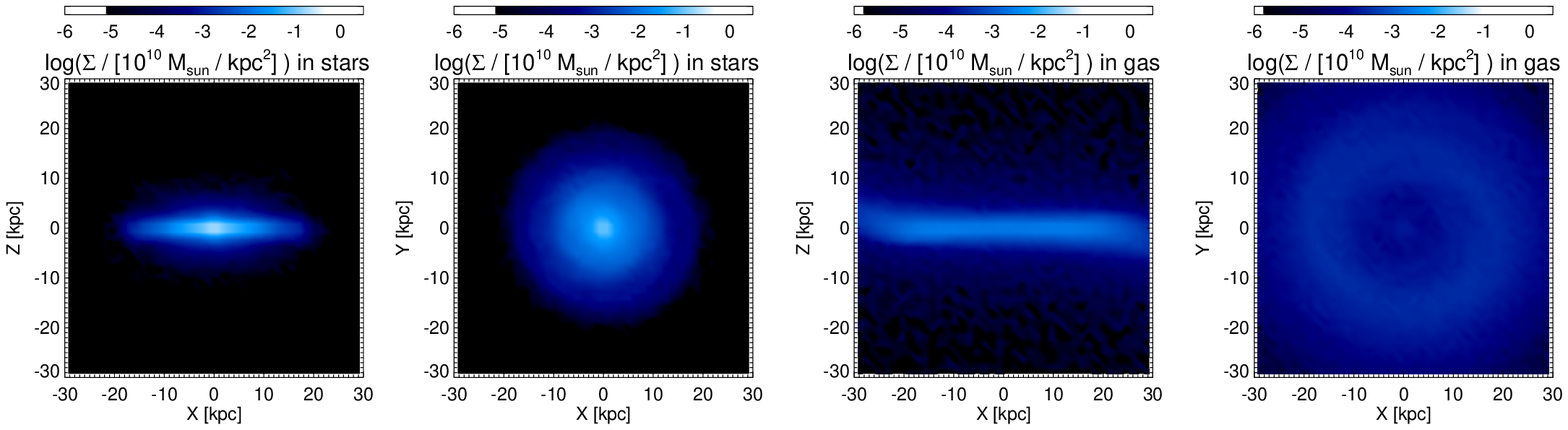}
\includegraphics[width=14. cm]{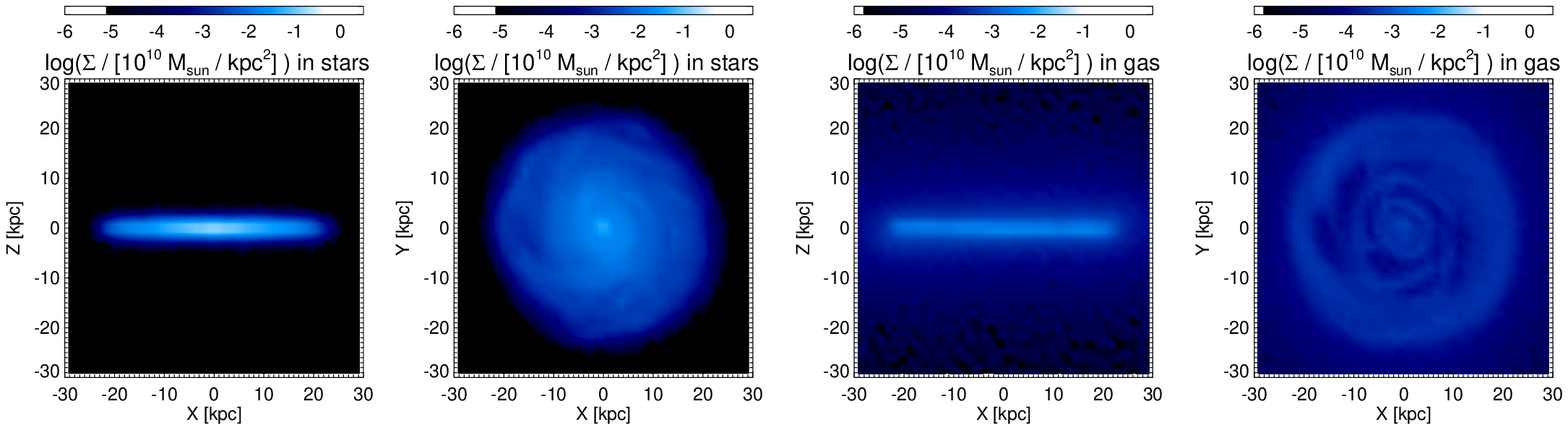}
\includegraphics[width=14. cm]{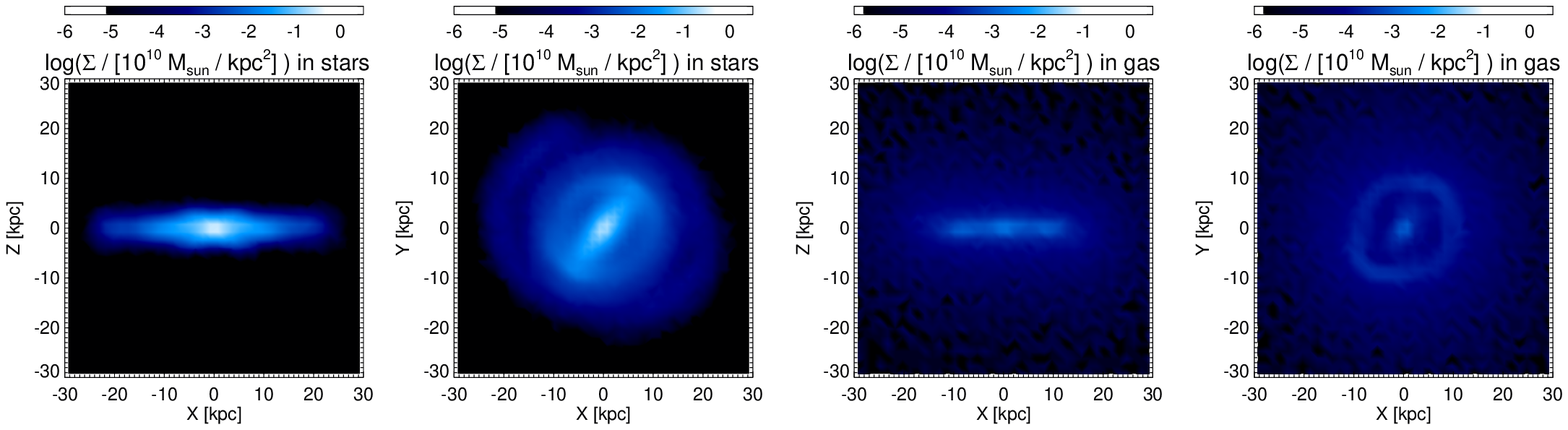}
\includegraphics[width=14. cm]{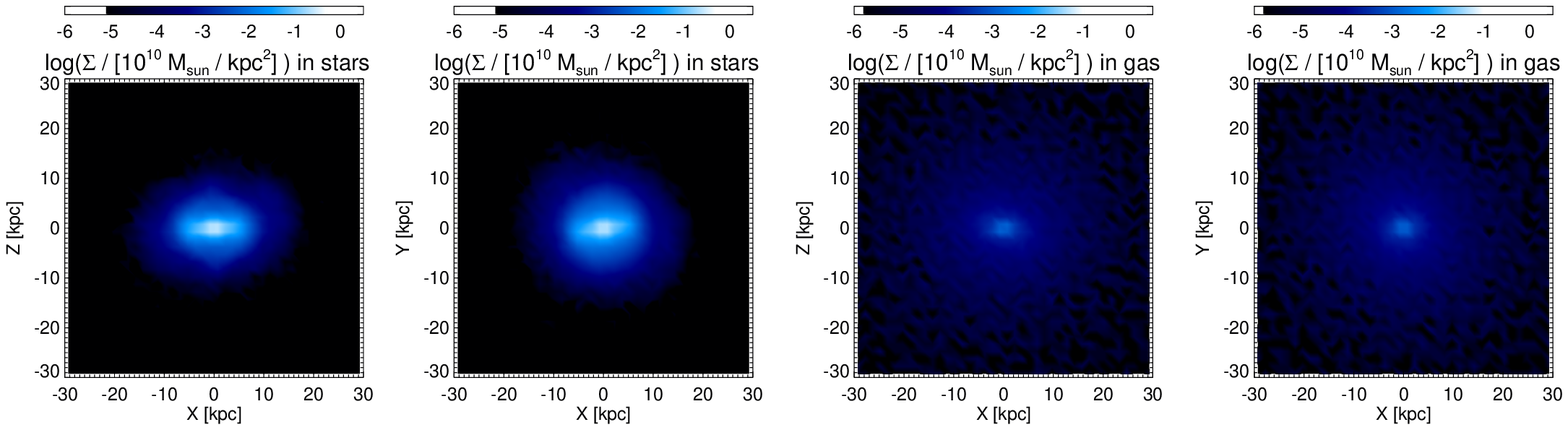}
\includegraphics[width=14. cm]{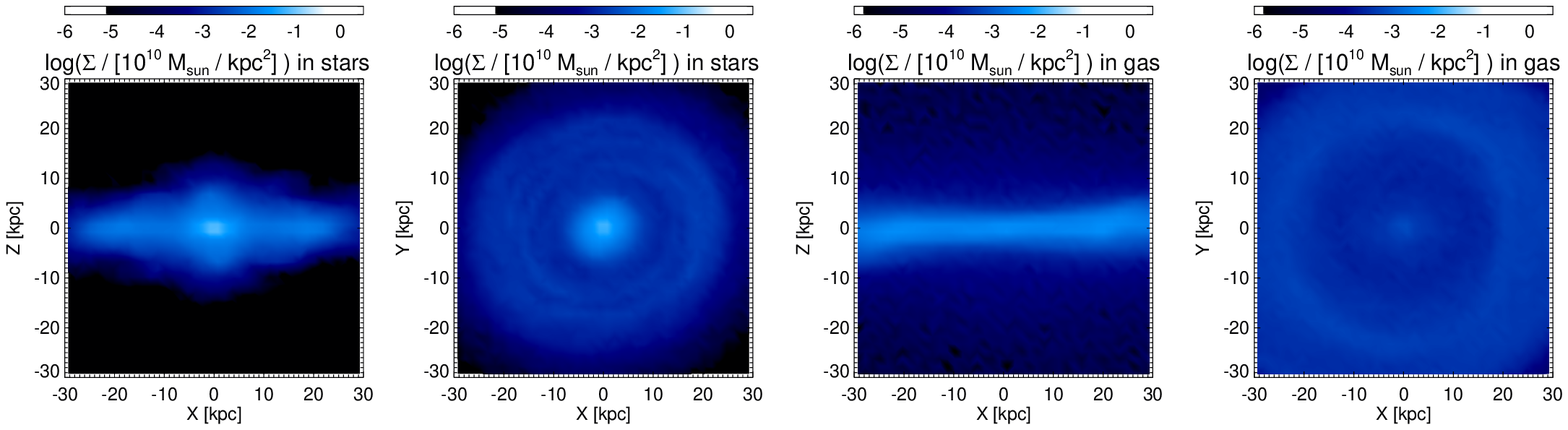}
\includegraphics[width=14. cm]{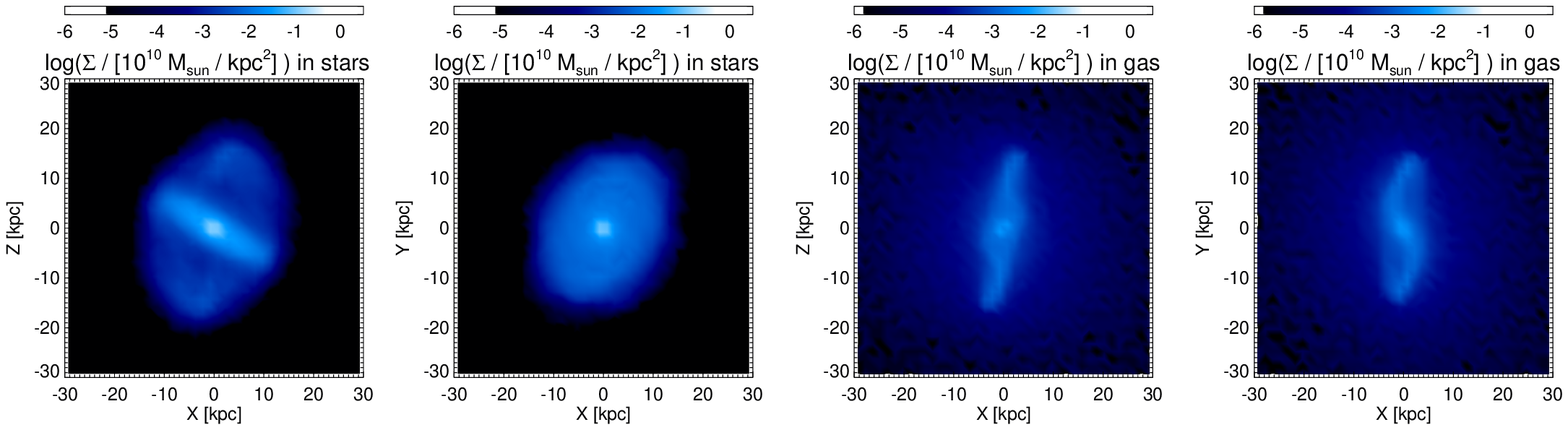}
\caption{Edge-on and face-on surface density projections for stars (left panels) and gas (right panels) at $z=0$ for models 
        (top to bottom): ARef, CExtCosmo, AExtLoAM, ARef-Min, CExt+AM and C-Maj.
}
\label{sd2d}
\end{figure*}

As halo A has an approximately prolate shape \citep{vera}, the axis ratio of the halo potential in the plane perpendicular to the major axis
is close to 1, i.e. almost axisymmetric, whereas it is $\sim 0.7$ if the disk axis is perpendicular to the major axis. Thus the formation of a disk
in the former case requires a less severe transition of the central halo potential into an oblate, axisymmetric, disk potential, than is the case
for the latter models. The strongest angular momentum losses in our models occur in these initial transformation phases. As our spherical initial
gas distributions are not in equilibrium with the gravitational potential
of the halo, the gas, which is significantly less dense than the
dark matter in the region in consideration, adjusts to the triaxial potential in the early phases of our simulations. Due to its rotation
the shape of the gas distribution becomes offset from the shape of the potential if the halo symmetry axis and the axis of rotation
do not agree. Gravitational torques then lead to the transfer of angular momentum to the halo. This explains the initial loss of angular
momentum in all models expect those oriented parallel to the major axis, as seen in figure \ref{angstuff}. The differences in the evolution of models
at the same orientation or with orientations perpendicular to major, which all show similar axis ratios of the equipotential ellipses in the plane
defined by their initial orientation, appear at later times (after $\sim 0.5-1$ Gyrs) and coincide with phases of reorientation of the angular 
momentum vector (see below).

In figure \ref{reorient}, we illustrate this reorientation of the total baryonic angular momentum vector in models ARef-Med, ARef+Med and ARef.
Strong reorientation starts to develop after $\sim 0.5-1$ Gyrs. At these times, the assembly of baryons in the centre of the halo has led to 
a transformation of the central halo potential shape, which apparently can induce strong reorientation of the baryonic angular momentum vector.
ARef shows the lowest amount of reorientation and has the lowest losses of angular momentum. ARef-Med shows continuous reorientation,
whereas ARef+Med settles into a new orientation relatively quickly. Consequently ARef-Med loses the most angular momentum
(see figure \ref{angstuff}).

In the right panel of figure \ref{angstuff} we focus on models with the same rotation axis. 
ARef180, which differs in orientation by 180 degrees from ARef, and
loses more angular momentum, as it shows strong reorientation. (figure \ref{reorient}). 
However, the losses are smaller than in ARef-Med, showing that angular momentum losses depend 
not only on reorientation and on the shape of the potential in the disk plane,
but also on the complex details of the interaction of hot gas, dark halo and stellar component.
AExt1C, which is more extended, more massive
and has a higher total angular momentum content, shows a smaller total angular momentum loss. AExt, which differs from AExt1C only by a 4 times
longer cooling time, loses approximately the same amount of angular momentum, but on a longer timescale, illustrating again that the loss
occurs as the gas cools to the centre. AExtLoAM, which has the lowest amount of angular momentum and goes bar unstable (see figure \ref{sd2d}),
shows smaller losses than AExt, which only differs in $V_{\rm{rot}}(R)$. 

For fixed values of other parameters, shorter cooling timescales, stronger reorientation
and higher initial densities lead to increased angular momentum loss. 
Higher initial angular momentum also leads to greater fractional losses.

\begin{figure}
\centering
\vspace{-.3cm}
\includegraphics[width=8.2cm]{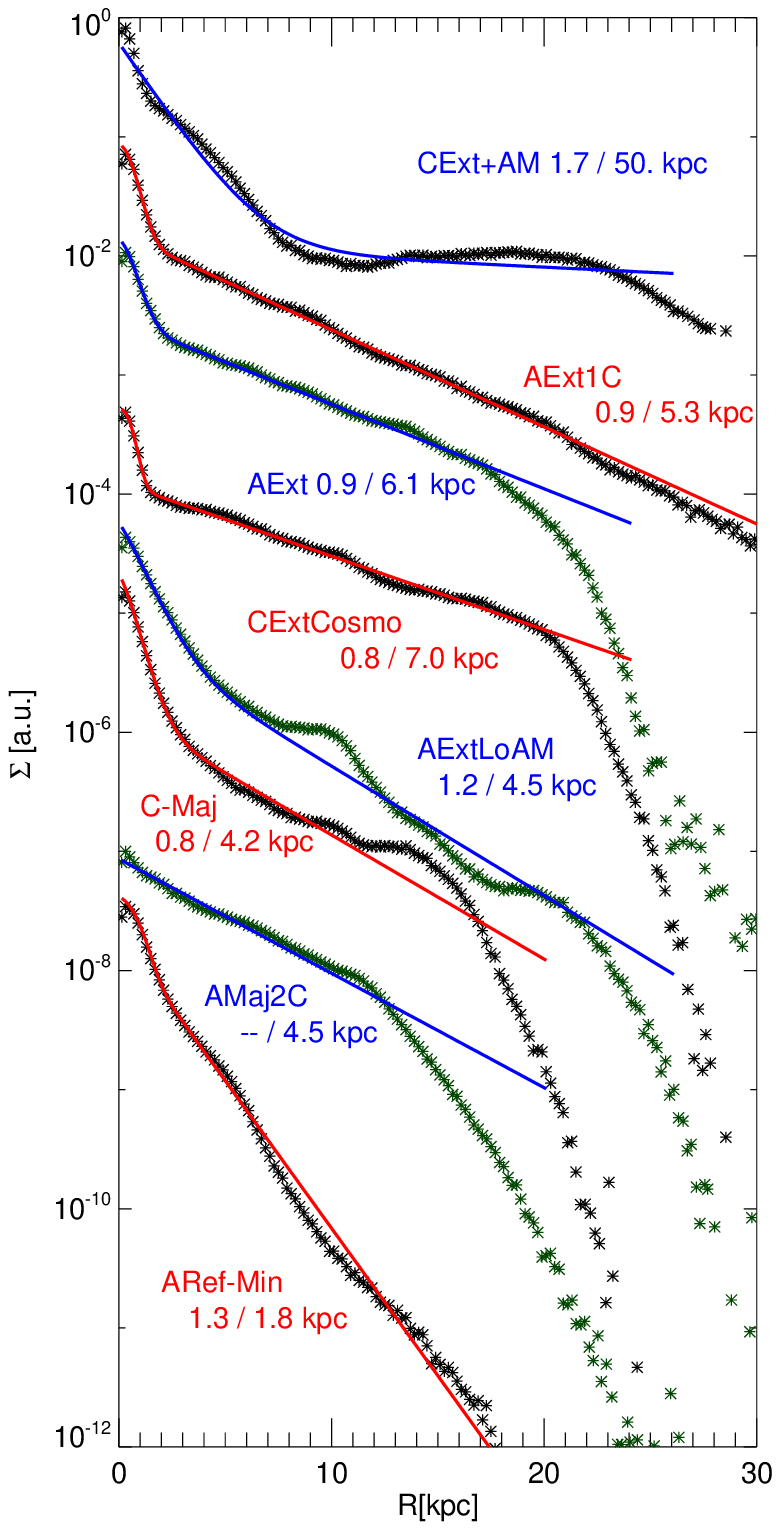}
\caption{Radial stellar surface density profiles, $\Sigma(R)$ at $z=0$ for models CExt+AM, AExt1C, AExt, CExtCosmo, AExtLoAM,
                               C-Maj, AMaj2C and ARef-Min. Overplotted are
                               fits of the type $ \Sigma_{0,1} \; \exp \left( - {{R}\over{R_d}} \right)
                               + \Sigma_{0,2} \; \exp \left[ - \left( {R} \over {R_b} \right)
                                 ^{\left({1} \over {n}\right)} \right] $. The numbers indicate the scalelengths
                               of the two components.}
\label{sfd}
\end{figure}

\subsection{Structural properties}

We continue with a study of the influence of the model parameters on the structure of the disk models.
In figure \ref{sfd}, we analyze $\Sigma(r)$ profiles for several models in halo A.
Here we focus on the models, that produce the strongest disks. The others are discussed in the following section.
Model AExt1C, which has more initial angular momentum, a higher mass and a larger 
initial radius than ARef (see bottom left panel in figure \ref{ARef})
produces a more extended disk with a disk-scale length $R_d=5.3 \;\rm{kpc}$
and no truncation out to 30 kpc. The four times longer cooling time of AExt mildly extends the scale-length to 6.1 kpc, but produces a truncation
around 20 kpc as the outermost material cools too slowly to form stars. 
The bulge profiles are very similar in the two models, showing that the formation of a centrally concentrated component due
to transformation of the central halo potential depends little on cooling time.

\begin{figure}
\centering
\includegraphics[width=8.5cm]{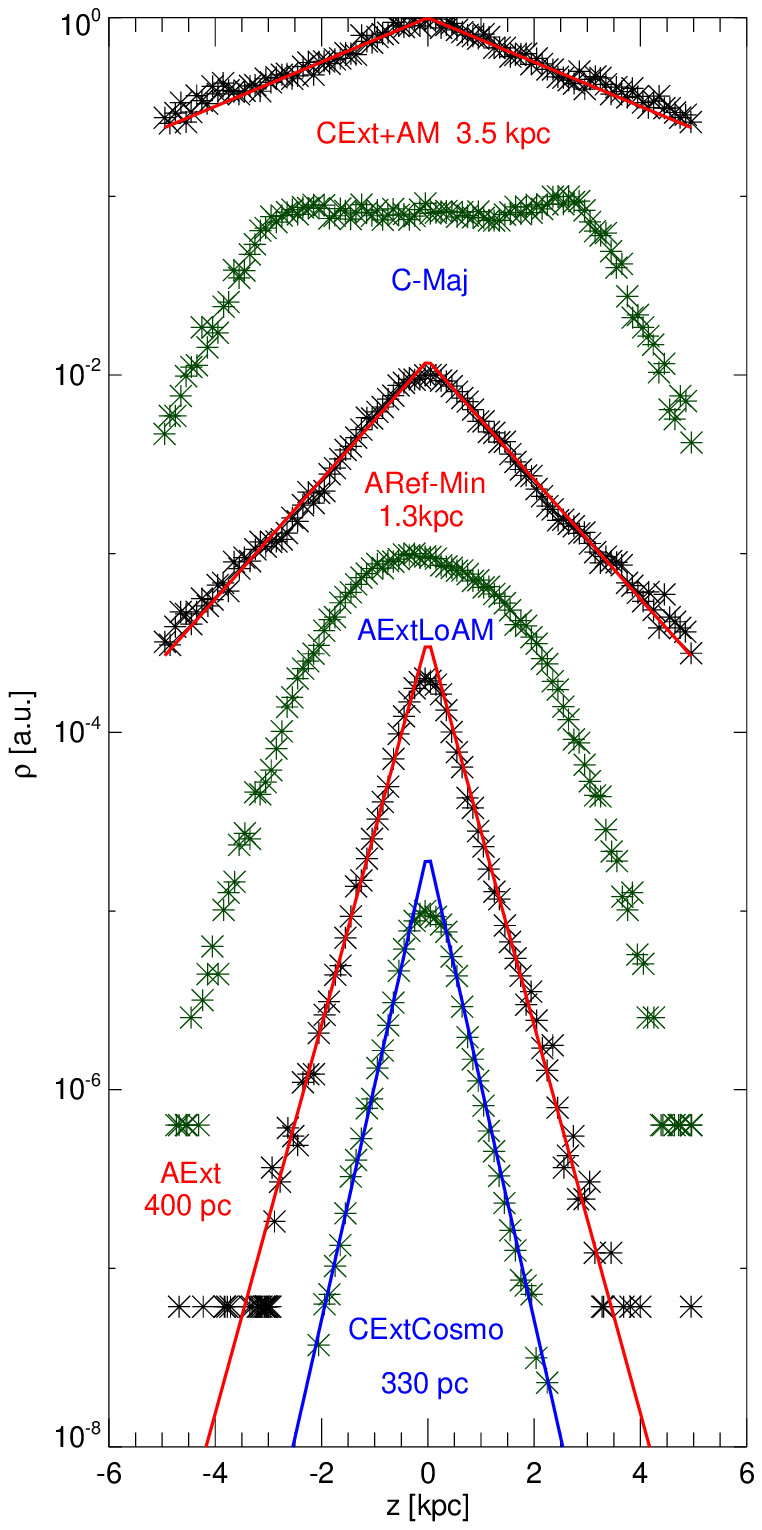}
\caption{Vertical density profiles of models CExt+AM, C-Maj, ARef-Min, AExtLoAM, AExt and CExtCosmo.
         If sensible, exponential fits are overplotted
         and the scaleheights are noted.}
\label{vert}
\end{figure}

Model AExtLoAM goes bar unstable and thus shows a much stronger central 
component with a longer scale-length (1.2 vs 0.9 kpc). The bulge profile extends to 8 kpc. The profile outside can be approximated by an exponential
disk, but clearly shows a ring structure, that is also visible in figure \ref{sd2d}, in both stars and gas.
Pseudo-bulge and ring are results of bar-induced gas flows as discussed e.g. in \citet{athan}.

In model AMaj2C there is no severe initial central transformation. The profile can be fitted by a single exponential, that,
because of the very low angular momentum losses, has a larger scale-length than ARef (4.5 vs 2.9 kpc)
despite the low initial angular momentum content. 
The low angular momentum also leads to a small truncation radius $R_{\rm{max}}\sim 12\; \rm {kpc}$.

The coldest disks form in AExt and CExtCosmo, which share initial density and angular momentum profiles as well as
cooling timescales, and which both undergo little reorientation. They show small halo-to-halo 
differences in terms of $\Sigma(R)$. AExt has a more massive
bulge component, a slightly less extended disk ($R_d=6.1$ versus $ 7.0\;\rm{kpc}$) and a slightly smaller truncation radius.
This underlines that the effects of the parameters discussed in this section are very similar in the two cases.

In figure \ref{vert} we compare vertical profiles of several models at $R=5\;\rm{kpc}$. 
In contrast to the double-exponential of ARef (see figure \ref{ARef}), AExt shows a single exponential profile with a slightly reduced exponential
scale-height $h_z =400 \; \rm{pc}$ (compared to 420 pc for ARef). We also present the profile of CExtCosmo.
It has an even thinner disk with $h_z=330 \; \rm{pc}$, consistent with the reduced bulge fraction discussed above.
Moreover, the flaring of these disks is significantly reduced compared to ARef.
Scale-heights for CExtCosmo are $h_z<400 \; \rm {pc}$ at $R<10 \;\rm{kpc}$ and  $h_z<550 \; \rm {pc}$ out to $R\sim 20 \;\rm{kpc}$.

The edge-on and face-on surface density projections in gas and stars of figure \ref{sd2d} (second row) nicely reveal that CExtCosmo sustains
spiral substructure in the stellar and gas disk until $z=0$, unlike ARef, which at this time has a significantly lower SFR.
A comparison to ARef also highlights the less prominent bulge structure and the lack of an underlying thick component.
A longer cooling time and a higher angular momentum content result in a prolonged star formation 
in the disk of CExtCosmo. Together with a small amount of reorientation of the
angular momentum vector, this leads to the continuous existence of a relatively thin stellar disk
preventing the formation of a thick component.

In comparison to these extended disks the low angular momentum model AExtLoAM, which goes bar unstable
shows a thick vertical profile at $R=5\;\rm{kpc}$ (in the bar region), which is fit neither by an exponential nor by an isothermal profile. 
In the edge-on view in the third row of figure \ref{sd2d}, the peanut-shape of the stellar system is clearly visible.

\begin{figure}
\centering
\includegraphics[width=8.5cm]{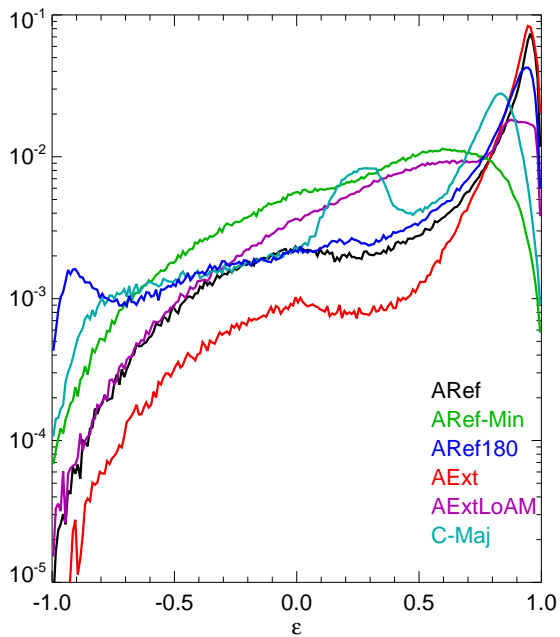}
\caption{Circularity distributions for models ARef, ARef-Min, AExt, AExtLoAM, ARef180 and C-Maj.}
\label{eps}
\end{figure}

\subsection{Kinematical Properties}

The significant parameter dependencies found above are also reflected in the stellar kinematics of our models.
In terms of the circularity distribution, longer cooling times and the resulting thinner and more extended disks
result in a slightly enhanced disk peak and a reduced bulge peak as may be seen by comparing AExt and ARef in figure \ref {eps}.
The massive bar in AExtLoAM significantly diminishes the $\epsilon \sim 1$ peak and shifts material into a wide, asymmetric distribution
at $-1<\epsilon<0.8$. 

\begin{figure}
\centering
\includegraphics[width=8.5 cm]{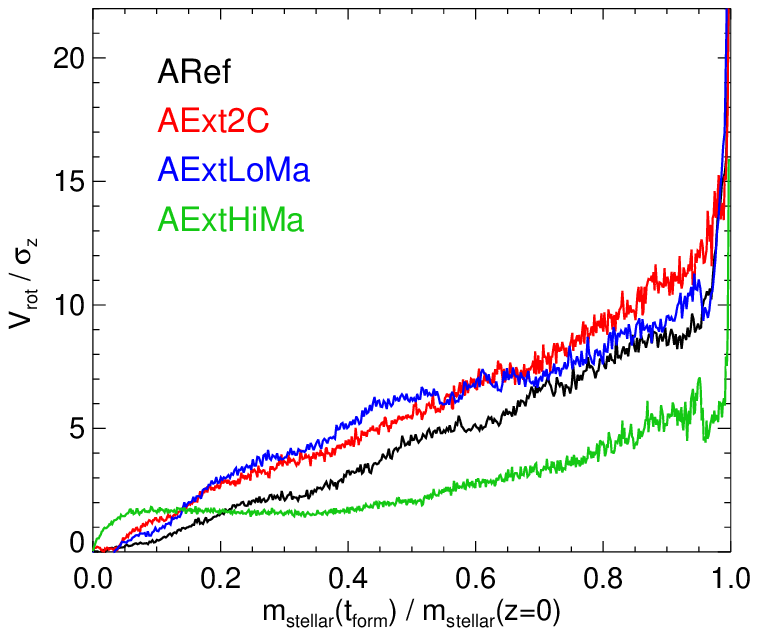}
\includegraphics[width=8.5 cm]{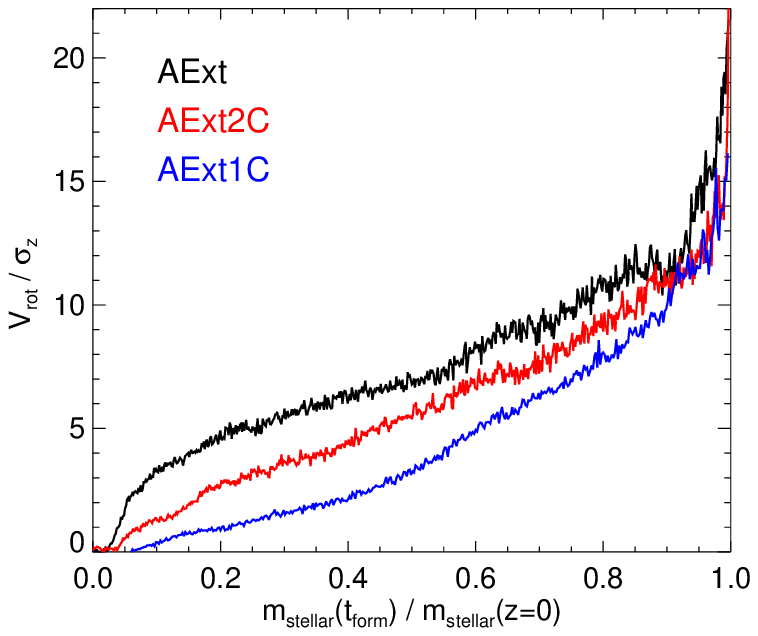}
\includegraphics[width=8.5 cm]{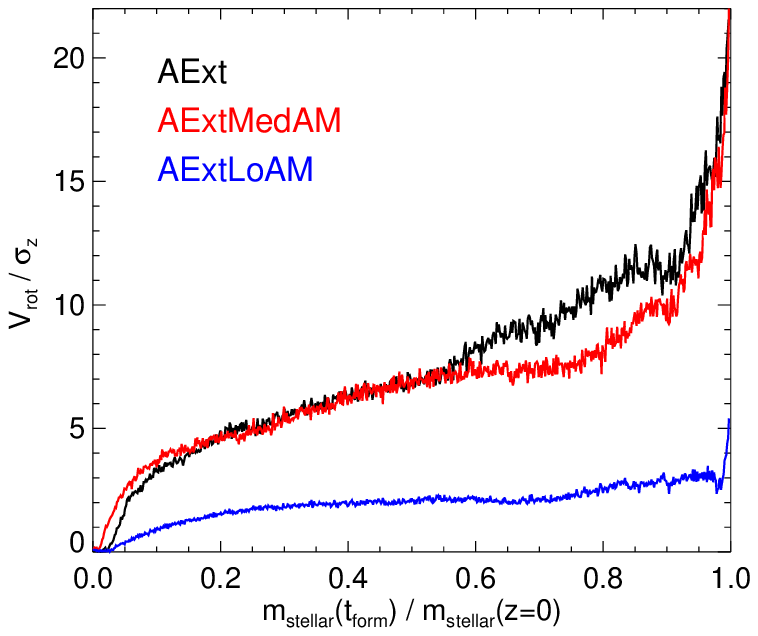}
\caption{Rotation-to-dispersion ratio as in figure \ref{vs1}, but now for different models at $z=0$.
         \textit{Top}: The effect of the initial radius $R_{\rm{gas}}$ and mass $M_{\rm{gas}}$ illustrated by models ARef, AExtLoMa, 
                             AExt2C and AExtHiMa.
         \textit{Middle}: The effect of the normalization of the cooling time $t_{\rm{cool,0}}$ illustrated by models AExt1C, AExt2C and AExt.
         \textit{Bottom}: The effect of the angular momentum content / rotation velocity profile $V_{\rm{rot}}(r)$
                                illustrated by models AExt, AExtMedAM and AExtLoAM.
}
\label{vs2}
\end{figure}

\begin{figure*}
\centering
\includegraphics[width=8.5cm]{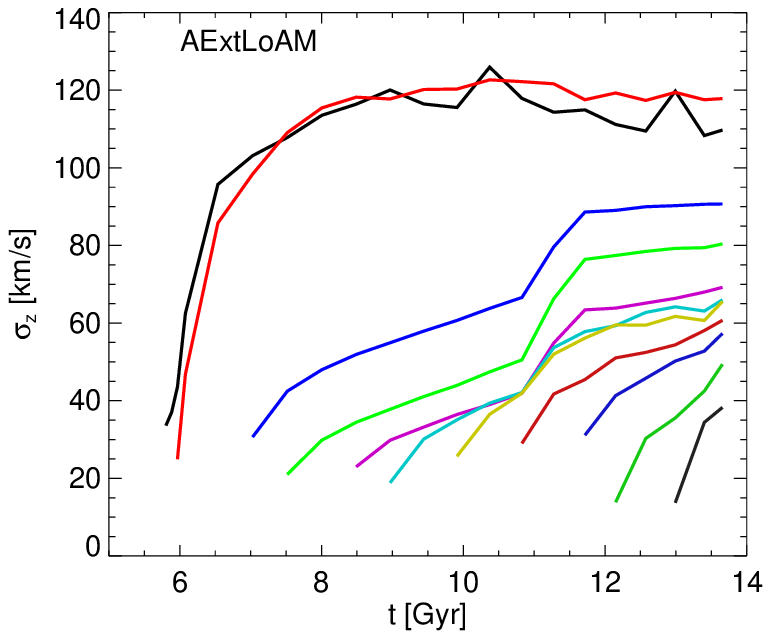}
\includegraphics[width=8.5cm]{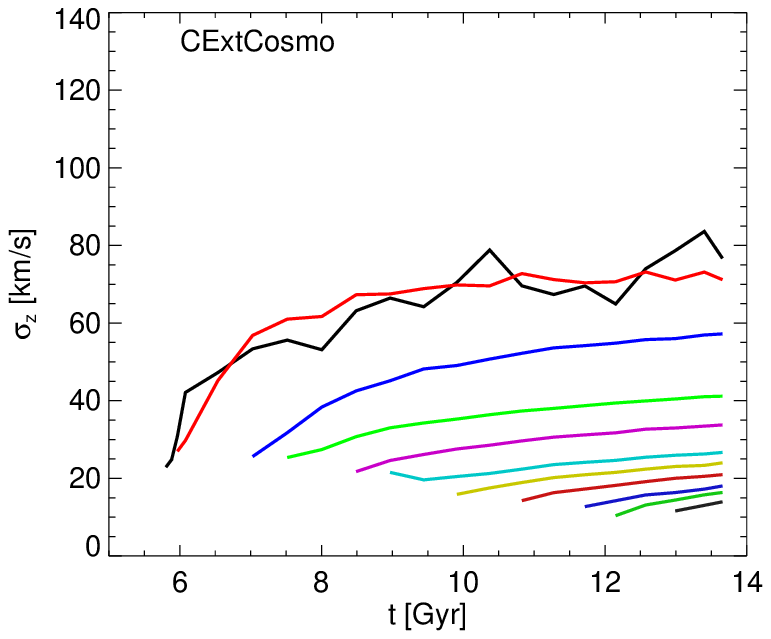}
\caption{Vertical velocity dispersions $\sigma_z(t)$ as a function of time for coeval populations in models 
         AExtLoAM (left) and CExtCosmo (right). See figure \ref{vs1} for an explanation of how
         these plots are constructed.
}
\label{diskheating}
\end{figure*}

The parameter dependence of the disks is summarized in the rotation-to-dispersion-ratio plots of
figure \ref{vs2}. In the upper panel we compare models ARef, AExtLoMa and AExt2C. 
Compared to ARef, AExtLoMa is more extended ($R_{\rm{gas}}=100$ vs 60 kpc).
This leads to an increase of $V_{\rm{rot}}/\sigma_z$ for all generations of stars, a reduction of the bulge-fraction, but no significant 
increase in the thin disk population. A lower density, which already results in slower cooling (cf. figure \ref{tff}), thus helps
reduce the impact of initial bulge formation. An increase in mass by 50\% has little impact as may be seen 
by comparing AExt2C to AExtLoMa.
It leads to more $z=0$ thin disk stars as it increases the surface density and thus prolongs star formation in the
disk outskirts. Increasing the mass again by a factor of 2, as in AExtHiMa, reduces the rotation-to-dispersion-ratio significantly.
The bulge fraction is reduced due to the longer cooling time applied (see middle panel). 
The disk goes bar unstable, however, and the bar subsequently heats the disk.

The second panel compares the models AExt1C, AExt2C and AExt and thus reveals the impact of the cooling time. Clearly a longer cooling 
time leads to an increase of $V_{\rm{rot}}/\sigma_z$ for all generations of stars. The increase is especially significant for the 
old population. AExt only shows $\sim 5\%$ bulge stars. From figure \ref{vs1} we know that $V_{\rm{rot}}/\sigma_z$ decreases due to disk heating 
as stellar populations age. As we know from figure \ref{sfr} that AExt has a much younger population than AExt1C, the difference
for the disk stars can be attributed mainly to the younger ages of the population. $V_{\rm{rot}}/\sigma_z$ for the thin disk population
is similar for all models, confirming that this is mainly determined by resolution and by the prescriptions for star formation and for cooling.

The bottom panel of figure \ref{vs2} concerns the initial angular momentum content of the models. Compared to AExt, $V_{\rm{rot}}(R)$
in AExtMedAM has been reduced by $1/3$ and by $2/3$ in AExtLoAM. Interestingly, AExtMedAM shows a slightly smaller dispersion-dominated component.
This indicates that the transfer of angular momentum from baryons to the dark halo plays a crucial role in the initial phase of bulge 
formation/transformation of the central potential. For the younger population, AExt is colder than AExtMedAM, as it is more extended
and thus has longer star formation timescales for the outer populations. AExtLoAM has been discussed above and shown to go massively bar unstable.
The consequence is a drastic reduction of $V_{\rm{rot}}/\sigma_z$ for all ages.

\subsection{Disk heating}

To illustrate differences in disk heating, we depict in figure \ref{diskheating} the evolution of $\sigma_z(t)$ for several coeval populations
of stars in models AExtLoAM and CExtCosmo.
Compared to model ARef, presented in figure \ref{vs1}, they both have lower initial densities and slower cooling times.
Like ARef they show continuous disk heating for all populations.
Moreover, the oldest populations undergo strong heating in the first several $10^8$ years, but, unlike ARef, the slower initial
transformation of the central potential prevents hot initial dispersions $\sigma_{z,\rm{ini}}>50\;\rm{kms^{-1}}$.

The low angular momentum model AExtLoAM shows the strongest heating of old components (to $\sigma_z\sim120\;\rm{kms^{-1}}$) due to the 
most concentrated surface density profile. At $t\sim 11 \;\rm {Gyr}$ after the massive bar has developed (cf. figure \ref{sd2d})
a strong enhancement in the heating of disk populations is visible. In less than a Gyr, $\sigma_z(t)$ increases by about 50\%
for all populations except the oldest, which are dispersion-dominated and are hardly affected by the bar.
An inspection of radial velocity dispersion $\sigma_R(t)$ reveals that in-plane heating is 
already enhanced at $t\sim 9 \;\rm {Gyr}$, when the bar starts to form. 
The bar also increases the birth velocity dispersions of stars by about 50\%. The final bar/disk populations are 
heated more efficiently and are thus
significantly hotter than in ARef. Analyzing $\sigma_z(t)$ for radial bins shows that the increase occurs at all radii less than 
$\sim 8\;\rm{kpc}$ and also in the central kpc. The stellar populations outside $R>\sim 10\;\rm{kpc}$
(outside the bar region) are however not affected.

For the disk populations in the high angular momentum model CExtCosmo, $\sigma_z(t)$ behaves as for the corresponding populations in ARef,
continuously increasing with the birth dispersions decreasing with time. The initial heating for the oldest populations is 
however significantly reduced compared to ARef. Consequently the transition between the oldest and the subsequent populations
is not as strong as in the other models. This is reflected by the increase in D/B for disk models with longer assembly timescales.
Still, CExtCosmo and all other disk models show a decrease in the birth velocity dispersions with time.
This behaviour is a consequence of the triaxial halo and its substructure, and does not occur in simulations within idealized spherical haloes.

\subsection{Summary}

In this section, we have studied the influence of cooling timescales, angular momentum content, initial density profile and mass on our models.
We show that the formation timescale increases with increasing cooling time and increasing angular momentum content. The angular momentum
loss in our models increases mainly with increasing initial ellipticity of the dark matter potential in the disk plane.
If reorientation of the baryonic angular momentum vector occurs in a model, it adds to the loss of angular momentum.
A slower formation process and thus a slower transformation of the central potential
yields thinner and more dominant final disks and weakens the formation of prominent bulge and thick disk components. Massive bulges can only be 
avoided if the initial potential contours in the disk plane are almost circular. Double-exponential vertical disk profiles are
suppressed by a slow formation timescale. All our disks show continuous secular heating due to (spiral) substructure (e.g. 
\citealp{jenkins}), so that younger disks are thinner. Disk heating is strongly enhanced by bars, as was recently also discussed by 
\citet{saha}. Additionally, heating is strongest for the old bulge populations and the birth velocity 
dispersion of stars in our simulations always decreases with time, which adds to the age-velocity
dispersion relation (see also \citealp{house}). Higher angular momentum content leads to more extended disks and larger truncation radii.
The latter also increase with faster cooling. Bars form, if, at fixed other parameters, the mass exceeds or the angular momentum falls below
(due to loss or initial lack) a stability limit consistent with 
standard stability criterions for bar formation (see \citealp{gd2} \S 6.3).
(Pseudo-)Bulges are enhanced by bar-induced gas inflows.

\begin{figure*}
\vspace{1.cm}
\centering
\includegraphics[width=17.cm]{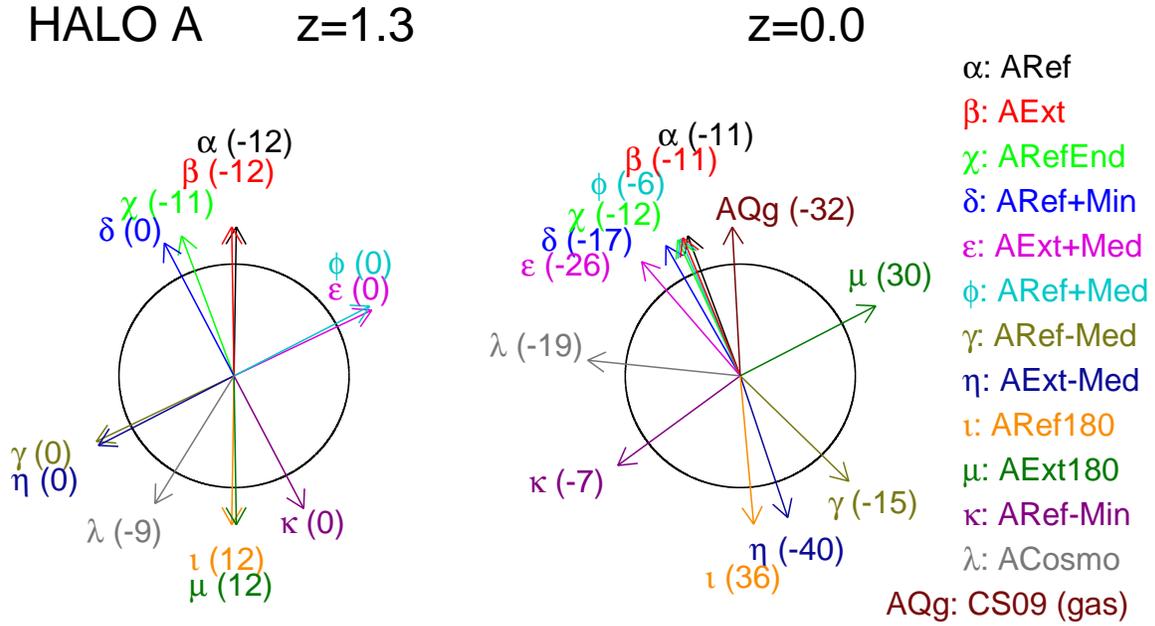}
\caption{Visualization of reorientation in halo A for models with initial orientations (approximately) in the plane perpendicular to the 
         major axis of the halo potential. Left are initial orientations at $z=1.3$, right are mass-weighted mean orientations of the stellar 
         systems at $z=0$. The coordinate system is static and was fixed to be perpendicular to the major axis of the halo potential at $z=1.3$
         and to have the initial orientation of model ARef at 0 degrees. The Greek letters and colors code the models
         and the numbers in brackets indicate the angle between the orientation and the plane in degrees. AQg stands for the orientation
         of the angular momentum of the gas at $R=30-100\;\rm{kpc}$ at $z=0$ in the simulation of CS09.
}
\label{arrowsA}
\end{figure*}

\begin{figure*}
\vspace{0cm}
\centering
\includegraphics[width=17.cm]{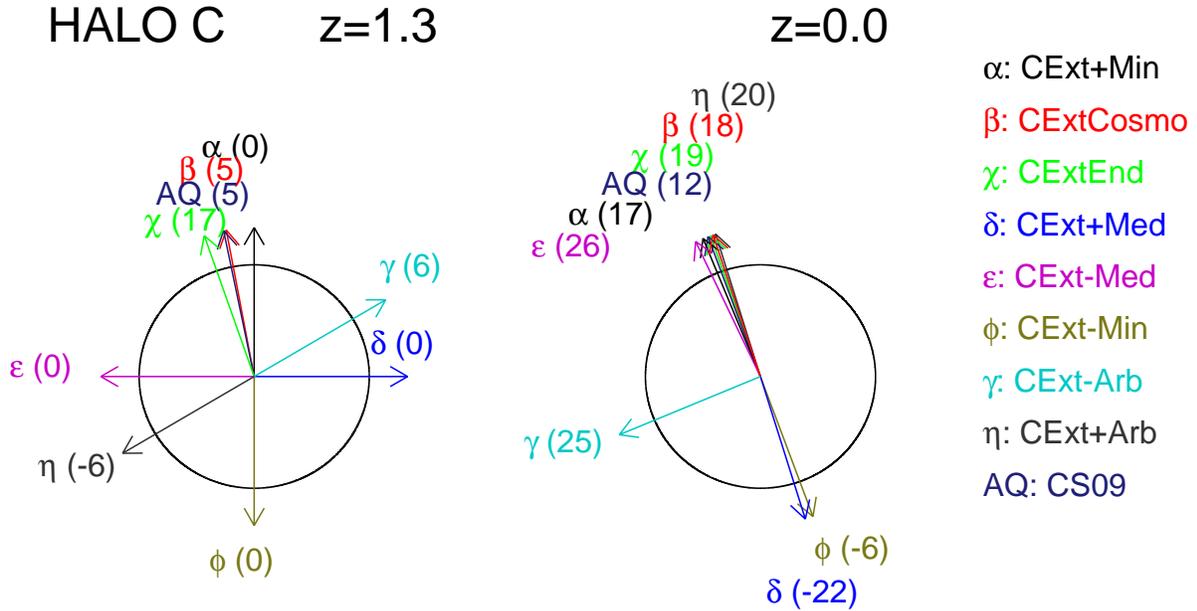}
\vspace{0cm}
\caption{As figure \ref{arrowsA} but for models in halo C. The initial orientation of model CExt+Min is at 0 degrees. AQ stands for the disk
         in halo C of CS09.}
\label{arrowsC}
\end{figure*}

\section{(Re)Orientations of disks}

So far we have paid relatively little attention to the influence of the initial orientation of the angular momentum vector of the rotating gas sphere.
We have shown that it plays a crucial role in angular momentum loss of the infalling gas (figure \ref{angstuff}). Moreover,
figure \ref{reorient} revealed that this orientation is not fixed, rather different models can undergo very different amounts of reorientation.
Model ARef and many others, which we used to study the influence of parameters, were initially oriented parallel to the angular momentum vector
of the dark matter within 300 kpc. This orientation was chosen as it yielded the coldest disks in several test simulations and showed
an amount of reorientation that is similar to that of the dark halo itself from $z=1.3$ to $z=0.0$. In terms of the principal
axes of the triaxial potential ellipsoid, the orientation of ARef is close to the minor axis, but produces slightly better results
than an exact alignment with this axis.

In halo C, this setup failed. The dark matter angular momentum is poorly aligned with the principal axes and thus the plane defined
by this orientation contains strong variations of the vertical gravitational force, preventing quiescent disk formation.
However, orientations close to the minor axis proved to work well, underlining that the influence of the shape of the halo potential is stronger
than the influence of its angular momentum. This finding is consistent with previous findings
that disks in cosmological simulations preferentially align with the minor axis of the halo \citep{bailin}.
In this section, we therefore explore which orientations in haloes A and C are capable of producing thin disks.
We do so in subsections, first discussing (re)orientations in minor and medium orientation models
and the effects of reorientation on kinematics. Then we discuss major/intermediate orientation models,
the structure of peculiar models and how reorientation shapes the potential.

\subsection{Models with orientations perpendicular to the major axis}

As both haloes are close to prolate at $z=1.3$ we start by exploring orientations, which lie in
or close to the plane perpendicular to the major axis of the halo potential. We illustrate the
findings for halo A in figure \ref{arrowsA}, where in the left half we show the initial 
orientations of the angular momentum of the gas sphere and in the right half we show the 
orientations of the angular momentum of all the stars that have formed by $z=0$. The coordinate 
system is fixed and identical in both halves. Zero degrees is defined by the initial orientation 
of ARef and the numbers in brackets behind the model numbers indicate the angles by which the 
orientations are offset from the plane in degrees. Initially all these models are within 12 
degrees of this plane. We show models aligned with the dark angular momentum within 300 kpc 
(ARef, AExt), 180 degrees offset from this (ARef180, AExt180), aligned with the minor axis 
(ARef+Min, ARef-Min) and with the medium axis (ARef+Med, AExt+Med, ARef-Med, AExt-Med) and 
aligned with the final orientation of model ARef (ARefEnd).

\begin{figure}
\centering
\vspace{-0.3cm}
\includegraphics[width=8.2cm]{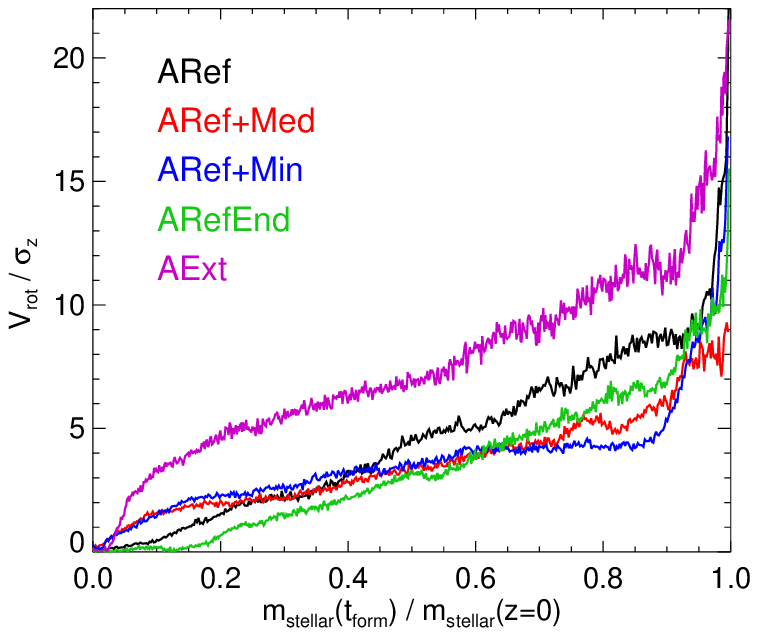}
\vspace{-0.3cm}
\includegraphics[width=8.2cm]{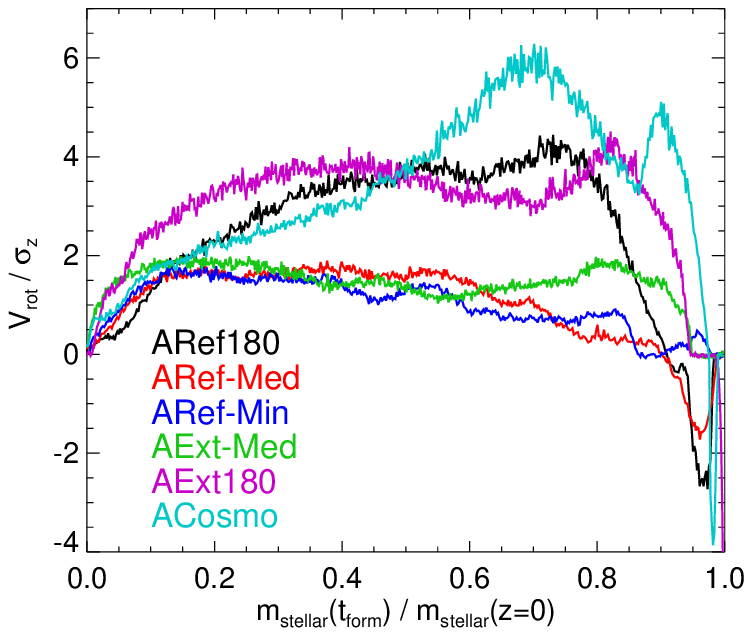}
\vspace{-0.3cm}
\includegraphics[width=8.2cm]{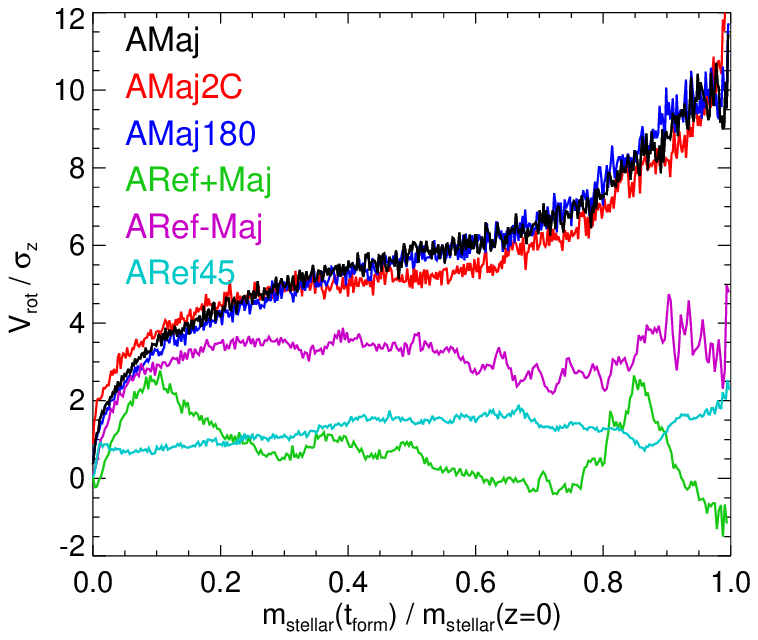}
\caption{Rotation-to dispersion ratios as defined in figure \ref{vs1} but now for models with different orientation.
         A negative ratio indicates a component that is counter-rotating with respect to the total stellar angular momentum.
         \textit{Top panel}: Models perpendicular to the major axis ending up in 
                             `disk orientation' and disk-like rotation: ARef+Med, ARef, ARef+Min, ARefEnd and AExt.
         \textit{Middle panel}: Models perpendicular to the major axis not showing a continuous 
                                disk formation, but all showing significant reorientation
                                and developing counter-rotating components: ARef-Med, ARef180, ARef-Min, AExt-Med, AExt180.
         \textit{Bottom panel}: Models parallel to the major axis: ARef+Maj, ARef-Maj, AMaj2C, AMaj, AMaj180
                                and model ARef45.
} 
\label{vs3}
\end{figure}

The final orientations of the galaxies in halo A show a clear crowding of models around the minor
axis. All models with orientations between ARef+Min and ARef+Med end up within $\sim 20$ degrees 
from the reference model ARef and close to the plane perpendicular to the initial potential major axis.
Figure \ref{reorient} shows a continuous reorientation of a total of 30 degrees for ARef. ARef+Med 
however first turns by $\sim 90$ degrees before also showing a small, continuous reorientation. 
Thus these models all seem to adjust to the `correct' orientation and then evolve rather quiescently. 
Models with different parameters sharing the same initial orientation show very similar final orientations
as illustrated by models ARef/AExt and ARef+Med/AExt+Med.
 
All the other models however end up with significantly different orientations, and there is no clear 
correlation between their initial and final orientation. ARef180 seems to end up close to its original 
orientation, however it flips by almost 120 degrees before slowly returning. If dark matter influenced 
the evolution only through its global potential, there should be no difference between two orientations
differing by 180 degrees, which is clearly not the case, so that halo rotation and transfer of angular 
momentum between the baryons and the dark matter must play a role in the described processes.
Comparing AExt180 and ARef180, which share initial orientation, we detect a final offset of more than 
90 degrees. A smaller, but significant difference exists also between AExt-Med and ARef-Med. In figure 
\ref{reorient}, one can see that the initial turn of ARef180 by $\sim 120$ degrees is also seen in 
AExt180, where this process takes longer. The model then stabilizes in this orientation. Model ARef-Med, 
which starts from a medium orientation, shows continual and strong reorientation.

In figure \ref{arrowsC}, we illustrate the situation in halo C. Zero degrees is defined by the initial 
orientation of CExt+Min, which was aligned with the halo minor axis. CExt-Min is 180 degrees offset, 
CExt-Med and CExt+Med are aligned with the medium axis and CExtEnd is the final orientation of CExt+Min.
CExt+Arb and CExt-Arb resulted from an error in the setup but are useful for this analysis, AQ marks 
the orientation of the disk in the cosmological galaxy formation simulations of CS09. Finally, CExtCosmo
uses the $z=1.3$ cosmological orientation as an initial condition. We see that the cosmological disk is
well aligned with the minor axis, around which many models again crowd at $z=0$, in this case all models
with orientations between those of CExt+Arb and CExt+Min in the initial conditions. CExt-Min and CExt+Med
indicate that, for halo C, the direction 180 degrees offset from the `best disk' orientation is also a 
preferred orientation at $z=0$. Among the models presented, only CExt-Arb ends up in a significantly 
different orientation.

\subsection{How reorientation affects kinematics}

In the previous subsection we have shown that the orientation of the angular momentum of infalling 
gas can undergo significant changes.
In the previous section we have already shown that this leads to loss of angular momentum. Here we study
the effects of these processes on the kinematics of the models.

In figure \ref{vs3} we show rotation-to-dispersion ratios for the halo A models ending up close to the ARef 
final orientation (top panel) and for the other models of figure \ref{arrowsA} (middle panel). In both 
panels, the models with longer cooling times show higher $V_{\rm{rot}}/\sigma_z$ for the disky populations 
as was discussed in the previous section. There is, however, a clear dichotomy between the two panels. 
All models in the upper panel show a monotonic increase in $V_{\rm{rot}}/\sigma_z$ from old to young stars, 
indicating that these are continuously forming disks. It also shows that ARef has the coldest disk 
population of the models sharing the same parameters, presumably because it also shows the smallest amount 
of reorientation. However, ARef+Min and ARef+Med, which start exactly in the plane perpendicular to the 
major axis show a smaller bulge population than ARef and ARefEnd, which start with a small misalignment 
to this plane.

The models depicted in the middle panel do not end up close to the preferred disk orientation and all 
show counter-rotating young populations indicating the infall of gas misaligned with the rotation of the
stellar object at late times and none of these objects show thin disk populations of stars.

Circularity distributions for ARef180 and ARef-Min are shown in figure \ref{eps}. The compact, barred disk
of ARef-Min has lost its $\epsilon \sim 1$ peak and shows a wide, asymmetric distribution peaking at 
$\epsilon \sim 0.6$. It has a strongly enhanced counter-rotating population compared to ARef. ARef180, 
in contrast, shows only a mildly diminished disk peak compared to ARef, but features a distinct, small 
counter-rotating disk peak at $\epsilon \sim (-1)$.

\begin{figure}
\centering
\includegraphics[width=8.5cm]{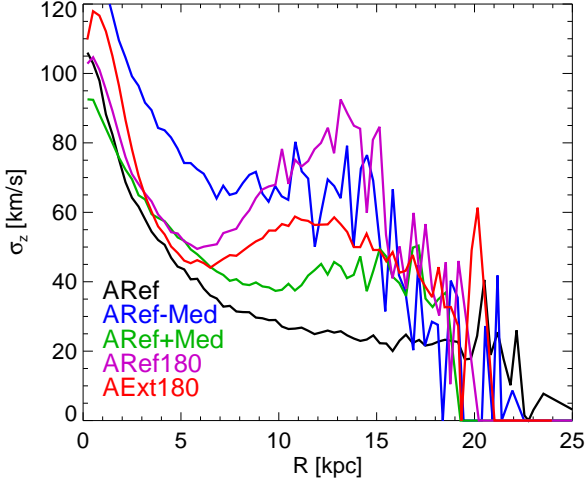}
\caption{Vertical velocity dispersions $\sigma_z(R)$ as a function of disk radius $R$ in the models ARef+Med, ARef, ARef180, ARef+Min, ARefEnd.
} 
\label{sigz}
\end{figure}

In figure \ref{sigz} we analyze the radial profiles of vertical velocity dispersion $\sigma_z(R)$ for 
several of the models just discussed. Unlike ARef, which shows a realistically declining $\sigma_z(R)$, 
there are several models showing an outward increase in $\sigma_z(R)$ after a drop at inner radii, most 
strikingly ARef180. According to figure \ref{reorient}, ARef180 shows the strongest reorientation during 
its evolution and according to figure \ref{angstuff} also the strongest losses of angular momentum of 
all models, whereas ARef shows only weak reorientation. This trend is confirmed by AExt180 and ARef+Med, 
which are intermediate between ARef and ARef180 in terms of $\sigma_z(R)$ flaring and reorientation.
ARef-Med, which shows the strongest angular momentum losses and continuously strong reorientation shows 
very high $\sigma_z(R)$ at all radii and mild flaring. This also strengthens the conclusion that these
two processes lead to high velocity dispersions in the final systems, especially in the outskirts.

\begin{figure}
\centering
\includegraphics[width=8.5cm]{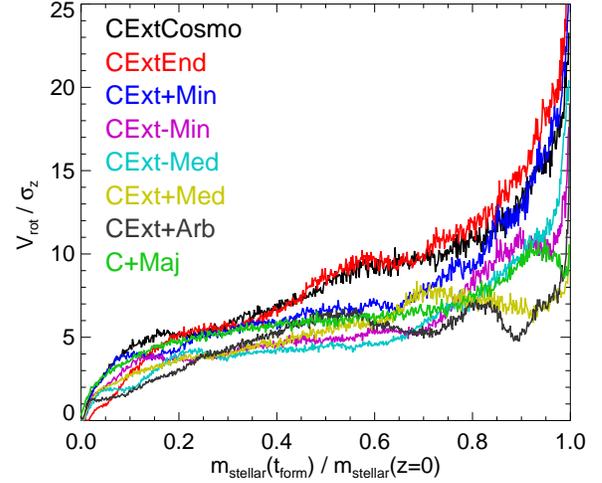}
\includegraphics[width=8.5cm]{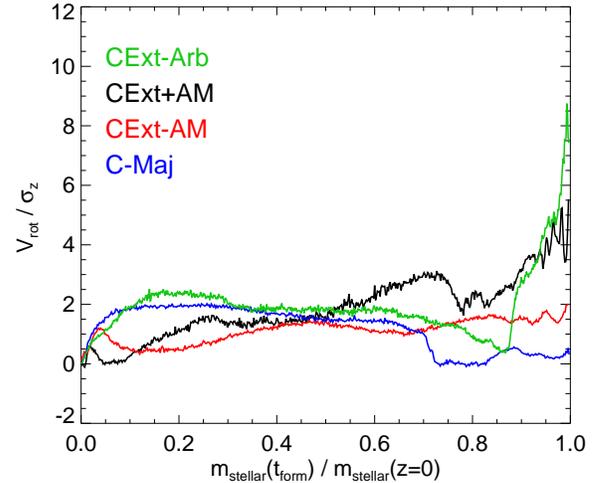}
\caption{Rotation-to dispersion ratios as defined in figure \ref{vs1}, but now for models in halo C.
         \textit{Top panel}: Models producing disks: CExt+Min, C+Maj, CExt-Min, CExt-Med, CExt+Med, CExtEnd, CExtCosmo, CExt+Arb.
         \textit{Bottom panel}: Other models in halo C: CExt+AM, CExt-AM, C-Maj, CExt-Arb.
} 
\label{vs4}
\end{figure}

In the top panel of figure \ref{vs4} we plot rotation-to-dispersion ratios for all models in halo C that
end up with a thin disk population of stars ($V/\sigma_z>5$ and $V/\sigma_z$ decreasing monotonically with
age). CExtEnd and CExtCosmo, the orientations motivated by the cosmological model and the final orientation
of CExt+Min, produce the coldest disks. Similarly to AExt in halo A, these models are slightly offset from
the minor orientation. CExt+Min, the minor orientation run, still has a better disk than CExt-Min, which 
starts and ends at an angle of 180 degrees. Unlike in halo A, this orientation still produces a disk.
The medium orientation runs CExt-Med and CExt+Med, which also start and end up at an angle of 180 degrees
produce slightly worse disks due to the effects of reorientation. CExt+Arb, which starts from an orientation
about 120 degrees offset from CExt+Min, also produces a disk in the preferred final direction. CExt-Arb is 
the only model not following the trend. The orientation changes strongly and the model results in a thick 
disk surrounded by misaligned components, each resulting from a different formation phase as indicated in 
the lower panel of figure \ref{vs4}.

\subsection{Model with major and intermediate initial orientations}

Until now, we have focused only on models that start (nearly) perpendicular to the major axis. 
As already discussed in the previous section, AMaj2C, which starts aligned with the major axis,
also produces a relatively small, thin and exponential disk. As can be seen in the bottom panel
of figure \ref{vs3}, the same is true for AMaj, which employs slower cooling, but produces only
a slightly colder disk. It was shown in figure \ref{sfr}, that the infall and star formation 
timescales are determined by angular momentum and cooling time. For these models with low angular 
momentum, star formation is shifted to earlier times, and consequently the stellar populations 
are subject to disk heating over a longer period. As a result the final difference in $\sigma_z$ 
is small.  AMaj180, which uses the same parameters as AMaj, but is initially 180 degrees offset, 
produces a disk with almost the same characteristics as AMaj. All these models reorient away 
from the initial major axis, but by different amounts, $\sim50$ degrees for AMaj2C and AMaj, 
which end only $\sim 5$ degrees offset, and $\sim 100$ degrees for AMaj180. They do not
end up near the minor axis, but do not seem to be affected much by reorientation. Probably this 
is because their formation timescales are small and they are compact. Misalignments typically 
happen for material that comes in late at large radii and is thus hardly affected by the potential
of the inner disk.

As alluded to in the previous section, the plane defined by the major axis orientation in almost
prolate systems has an almost axisymmetric potential and is thus advantageous for our models 
compared to the minor axis orientation. However, models ARef+Maj and ARef-Maj, which start with
significantly more angular momentum than AMaj/AMaj180 do not produce disk-like objects. This is 
also true for ARef45, which starts from an unfavorable orientation half-way between minor and major
axes, as can be seen in the bottom panel of figure \ref{vs3}. The reason for this failure is 
misalignments that develop between gas infall at early and late times, destroying an initial disk.

In halo C, the situation is similar. C+Maj, which starts in major axis orientation with the same 
parameters as AMaj, also produces a disk. It also undergoes reorientation away from the major axis
and ends up at an angle $\sim 30$ degrees from the final orientation of CExt+Min. However, C-Maj,
which has an initial gas angular momentum orientation 180 degrees apart from C+Maj and a more 
extended initial radius and thus higher angular momentum, does not produce a nice disk, showing
outer misalignments as in ARef+Maj. This model shows that misalignments are common for initial 
orientations parallel to the major halo axis, also for medium angular momentum content. 

The model CExt+AM is initially aligned with the angular momentum of the halo at $R<300\;\rm{kpc}$,
which is at an angle of $\sim45$ degrees to the major axis. This is why the model experiences 
similar problems to ARef45. The same is true for CExt-AM, which starts at an angle of 180 degrees
to CExt+AM.

\subsection{The structure of peculiar models}

Although our models were designed to produce disks, not all of them end up as such, as we have shown above.
Here we briefly analyze the structure of three of those models in more detail.

ARef-Min is one of the models, which start oriented perpendicular to the major axis and show continuous, 
strong reorientation, which leads to strong angular 
momentum losses and high velocity dispersion (see middle panel in figure \ref{vs3}). We have included surface
density projections of ARef-Min in figure \ref{sd2d} (fourth row). Due to the angular momentum losses
the resulting disk is compact, barred and thick with an exponential scale height of $h_z=1.3 \;\rm{kpc}$
(see figure \ref{vert}). Its exponential-like radial profile is very steep with $R_d \sim 1.8 \;\rm{kpc}$
as shown in figure \ref{sfd}.
The galaxy has a high bulge fraction, $D/B \sim 1$. Even at $z=0$ the remaining gas is concentrated in the 
central bulge and does not show a disk structure. Also the stellar disk structure is mostly bar-like and 
is surrounded by a spheroidal star distribution. Its circularity distribution (figure \ref{eps}) shows a 
wide, asymmetric peak at $\epsilon \sim 0.6$. The system might be classified as a barred, fast rotating
early-type galaxy applying a classification scheme as used for the ATLAS$^{3D}$ survey \citep{atlas}.

CExt+AM represents the models starting with intermediate initial orientations. In the fifth row of figure \ref{sd2d},
we show its $z=0$ surface density projections. The gas lives in a perfect disk, whereas the stellar disk shows two regions:
a dominant bulge-like central structure, which is elongated perpendicular to the gas disk, and outer regions dominated by
rings and separated from the centre by a drop in surface density around $R\sim 9\;\rm{kpc}$ (see figure \ref{sfd}). 
The vertical profile at $R=5\;\rm{kpc}$ can be fit by a thick exponential with $h_z\sim 3.5\;\rm{kpc}$ (figure \ref{vert}). 
Figure \ref{vs4} reveals several dips and peaks in $V_{\rm{rot}}/\sigma_z$ during the formation of the object indicating 
several misaligned populations, which prevent the formation of a cold disk. The system bears some resemblance with
polar bulge galaxies (see e.g. \citealp{4698}).

Models starting in major orientations and developing strong misalignments are represented by C-Maj, the surface
density projections of which are shown in the bottom row of figure \ref{sd2d}.  Two stellar components at an angle
$\sim 60$ degrees are clearly visible. Consequently, the circularity distribution in figure \ref{eps} shows two distinct
peaks at $\epsilon\sim 0.85$ and $\sim 0.25$. The latter represents the ring which forms at late times (see figure \ref{vs4}).
In the gas surface density projections, flows are visible, which channel the material to the centre. The vertical profile
at $R=5\;\rm{kpc}$ (figure \ref{vert}) is flat for $|z|<3.7\;\rm{kpc}$ before dropping exponentially, whereas the radial
profile (figure \ref{sfd}) resembles a double-exponential with a dominant central component at $R<5\;\rm{kpc}$
and rings at $R>10 \;\rm{kpc}$. These properties of the model reveal a similarity to polar ring galaxies \citep{whitmore}.

\subsection{How reorientation affects the shape of the potential}

In the context of the results already presented in this section, 
the question arises, how the shape of the final gravitational potential differs between models.
Clearly, in central regions dominated by a disk, the potential is found to be oblate. Due to the interaction with the
forming disk, the dark matter halo becomes less triaxial, in agreement with the results of \citet{kazantzidis}.
Moreover, the initial triaxiality of the halo, which in the dark matter only simulations shows little
reorientation, is still imprinted at outer, halo-dominated radii. Consequently, for models such as AExt or CExtCosmo, the gas angular
momentum of which was initially aligned with the minor potential axis, the final disk axis is well aligned with the final minor axis of
the potential at all radii. For disk models, which end up in preferred disk orientation, but started with a different
initial orientation, such as ARef+Med or CExt+Arb, the situation is similar, yet the potential contours are less flattened, 
consistent with the fact that these disks are thicker. For models, where the final stellar angular momentum does not align with the 
preferred disk orientation, such as ARef-Med or CExt-Arb, there is also no alignment between the final potential ellipsoids at inner
and outer radii. This is also true for compact disk models, that formed from gas which was initially aligned with the major axis.
The disks, which are finally closest to an alignment with the major axis of the outer potential are AMaj and AMaj2C with
an offset angle of $\sim 30$ degrees.

\begin{figure*}
\centering
\includegraphics[width=7.5cm]{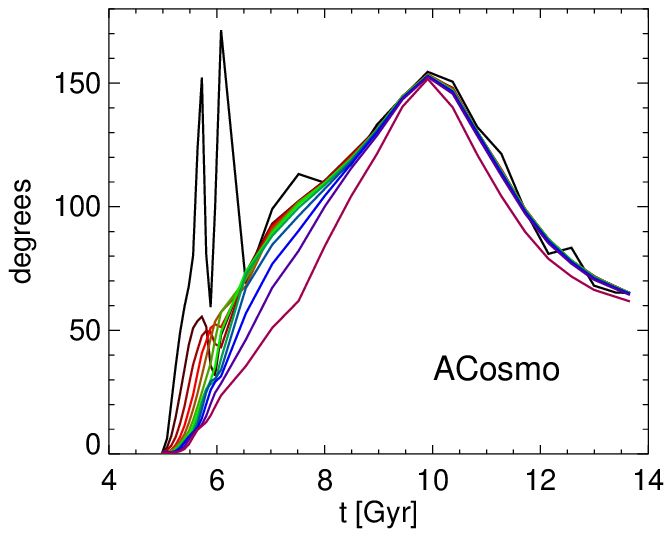}
\includegraphics[width=7.5cm]{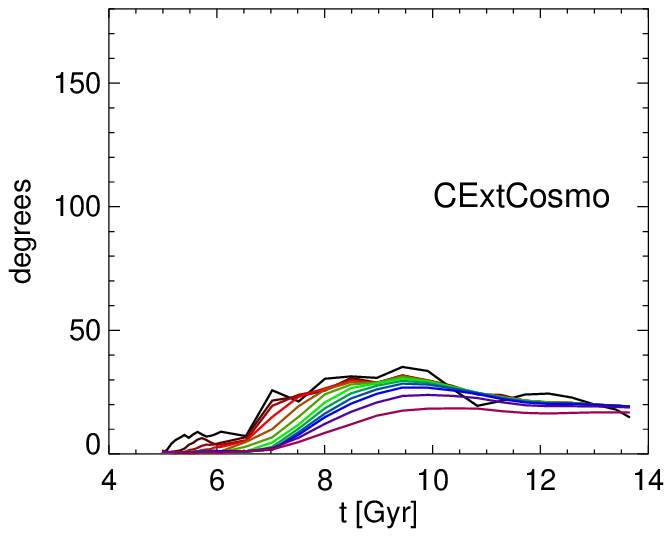}
\includegraphics[width=7.5cm]{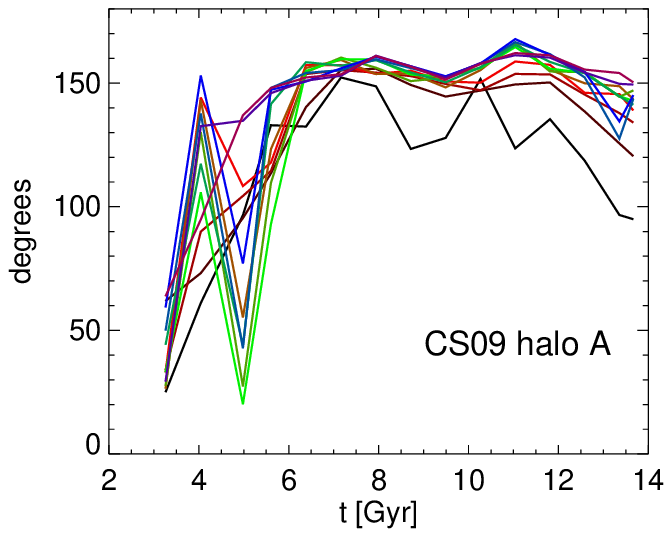}
\includegraphics[width=7.5cm]{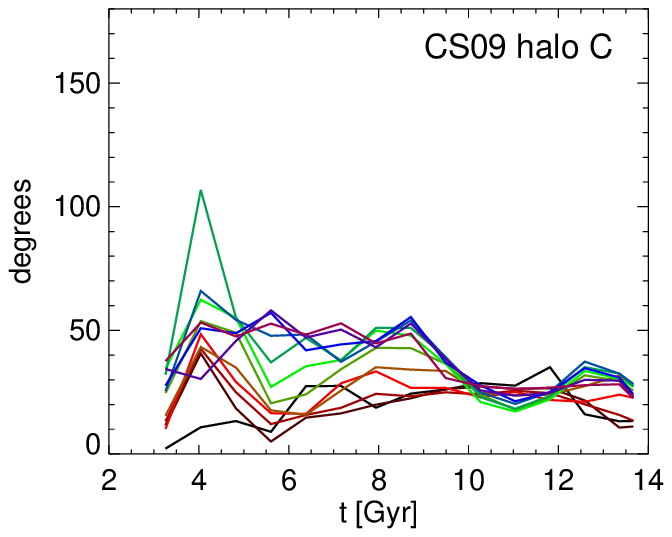}
\caption{Orientation angle as a function of time for models ACosmo (top left)and CExtCosmo (top right)
         The initial gas sphere has been divided into 12 radial shells of $R_{\rm{gas}}/12$ width each. 
         For each shell the angle of the angular momentum vector $\bf{j}_i$(t) of all particles
         initially in shell \textit{i} to the initial orientation is calculated at each output time $t$.
         The panels in the bottom row show the spin orientation in the same coordinate system for
         the gas content of 12 spatially \textit{fixed} concentric shells of 10 kpc width
         for the A and C runs of CS09.}
\label{cang2}
\end{figure*}

\subsection{Summary}

In conclusion, we have shown, that each of the two haloes we studied shows a preferred orientation for disks, which roughly agrees
with the minor axis of the inner halo potential, in agreement with the results of \citet{bailin} for fully
cosmological simulations. However a flip of the initial angular momentum vector by 180 degrees produces a slightly worse (halo C) or
unsuitable (halo A) orientation for a forming disk. This shows that it is not the shape of the potential alone that determines the preferred
orientation. Models with initial non-preferred angular momentum orientations tend to reorient to a
preferred axis, producing thickened and flaring disks. In some cases they fail to settle to a stable orientation and do not end 
up with significant disks. There is no obvious simple criterion to explain this dichotomy. Models starting with angular momentum
parallel to the major halo axis also do not show a stable orientation. They form stable disks only if their angular momentum content
is low, so that they rapidly form compact objects, which cannot be destroyed by reorientation, which misaligns inner and outer components. 
Realistic thin disks with continuously forming populations are thus only possible for initial orientations close to the halo minor axis.
Models with initial orientations in between these cases are not capable of producing cold disks.
Taking results of the previous sections into account, we have 
also shown that the orientations both of the central disk and of the outer gas can change by more than 90 degrees,
despite the simple and coherent initial distributions of our gas components.
This is also true for the angle between the components.

\begin{figure*}
\centering
\includegraphics[width=7.5cm]{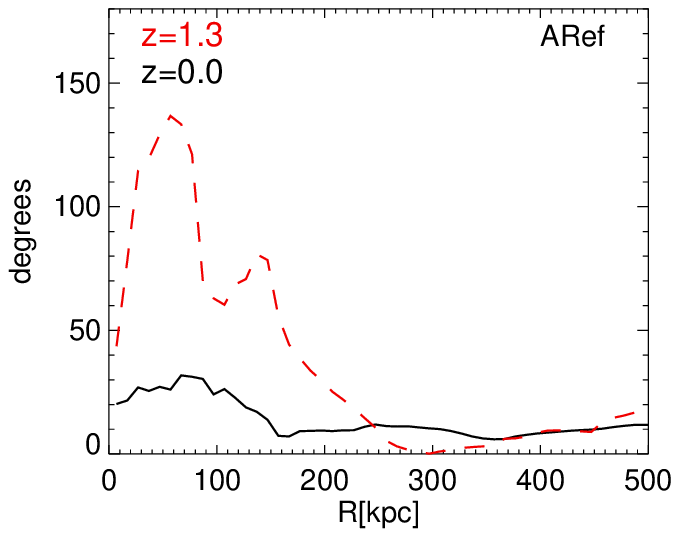}
\includegraphics[width=7.5cm]{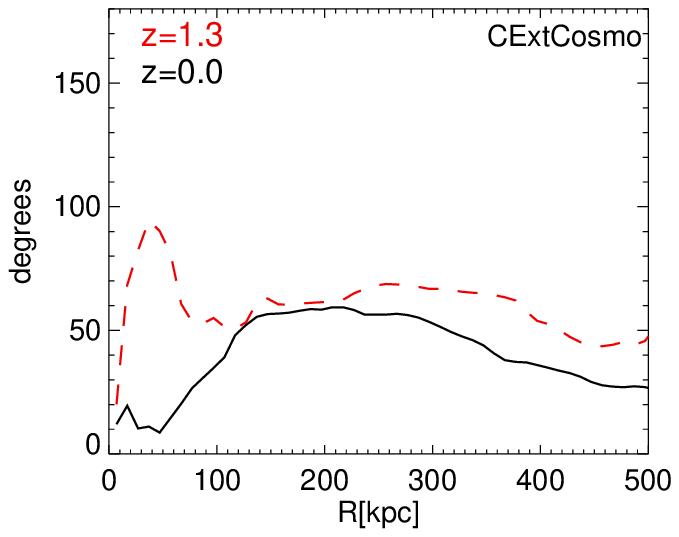}
\includegraphics[width=7.5cm]{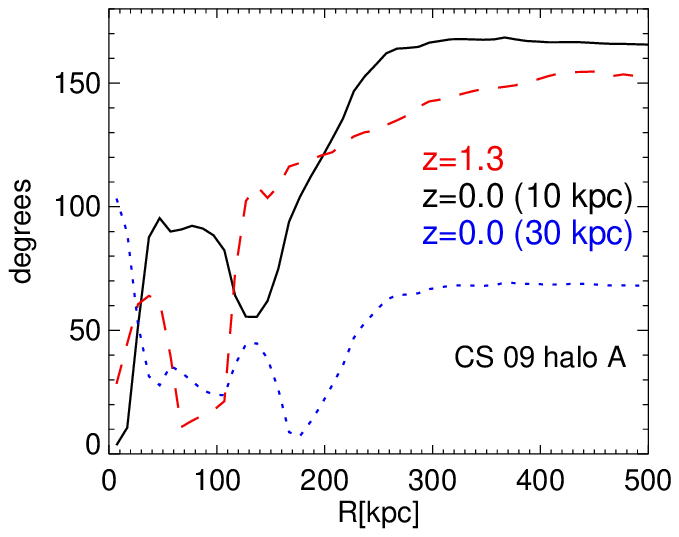}
\includegraphics[width=7.5cm]{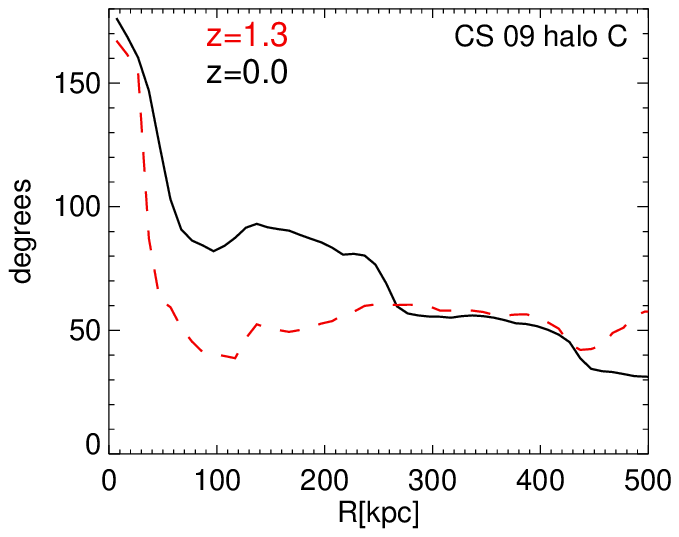}
\caption{The angle between the angular momentum vector of all stars within $R<10\;\rm{kpc}$ and the angular momentum vector
         of all dark matter within R as a function of R at $z=1.3$ and $z=0.0$ for models ARef (top left), CExtCosmo (top right)
         and the results of CS09 for haloes A (bottom left) and C (bottom right). The effect of varying the radius that
         defines the galaxy for halo A is indicated in the bottom left panel, where the blue line is for a 30 kpc galaxy radius.
         For all other models this radius definition plays a minor role.
}
\label{align}
\end{figure*}

\section{Comparison to cosmological simulations}

Until now, we have shown that, with our method, we are capable of introducing disks into Aquarius haloes A and C.
Using the cosmological orientations of the disks at $z=1.3$ in the simulations of CS09, we showed that for halo
C the disk is in the `right place', meaning that our model CExtCosmo, which starts with this orientation, produces a nice disk
at $z=0$. In halo A, the cosmological disk is in the `wrong place', i.e. our model ACosmo, starting in the corresponding orientation,
fails to produce a nice disk. We have identified reorientation of the angular momentum of infalling material and the angular momentum losses
and misalignments connected to this phenomenon as the key factors in the failure of many of our models to build substantial disks. CS09 also
identified misaligned infall of gas as the problem destroying the existing disk structure in their resimulation of halo A.

In this section we would therefore like to analyze the reorientation of gas in the runs of CS09 and compare them
to our own. In these cosmological simulations the inflow of gas is continuous and more complex than in ours, it is not desirable to define fixed
gas populations by their radial position at a certain time. It is more instructive to follow how the angular momentum
of gas evolves in fixed radial shells of 10 kpc width as is shown in figure \ref{cang2}, where we compare the reorientation in
models ACosmo and CExtCosmo to the reorientation in models A and C from CS09.

ACosmo, as depicted in the top left panel of figure \ref{cang2}, shows strong and continuous 
reorientation for all baryons, gaseous and stellar. Reorientation is slightly
delayed for the outer shells compared to the inner, but the resulting misalignments are small compared to the models discussed 
in the previous section. After $t\sim9\;\rm{Gyr}$ there is no overall detectable misalignment among the initially defined shells, although
there is a small counter-rotating component as shown in figure \ref{vs3} originating from temporary misalignments and
continuous reorientation. As shown in figure \ref{arrowsA}, ACosmo is $\sim 45$ degrees offset from the orientation 
of most of the disk models in halo A at $z=0$. During its evolution it evolves towards this orientation until $t\sim 11 \;\rm{Gyr}$,
when it is only $\sim 20$ degrees offset and then starts to turn away again.

For the fully cosmological model we start the analysis at $z=2$ (lower left panel). The reference orientation is the one of the
stars within $R<10\;\rm{kpc}$ at $z=1.3$, which is the initial orientation of
ACosmo. At $z=2$ the orientation of the stellar angular momentum differs by less than 10 degrees.
Even at $z=2$ the gas out to $R=120\;\rm{kpc}$ is not well aligned with the galaxy. Until $z\sim1$ there is a clear misaligning
process. From then on, gas in all shells shows an orientation $\sim 150$ degrees offset from that of the galaxy at $z=1.3$. 
The new orientation is also depicted in figure \ref{arrowsA}. It only differs by $\sim 20$ degrees from the initial orientation of model
ARef, which explains why this orientation yielded the best results. The offset is caused by a stronger misalignment with the plane
perpendicular to the major axis.

The strong reorientation of the gas is also depicted in figure 10 of CS09, where the evolution of the angle between the stellar and the 
gas components within 27 comoving kpc is plotted. 
The fact that the components become aligned again at $z=0$ indicates that the inner stellar component reorients
to the same direction, but with a delay. Moreover, the central gas components (black lines in figure \ref{cang2}) are also
offset from the outer gas by up to $\sim 50$ degrees.

The new orientation of the (outer) gas is also within $\sim20$ degrees of the orientation that model ACosmo shows at $t\sim 10.5 \;\rm{Gyr}$, but
at $z=0$ the orientations differ by $\sim 70$ degrees. This is due to the continual reorientation in model ACosmo.
While the reorientation process is qualitatively similar in these models, the final orientation of the galaxy is
also dependent on the details of its assembly, as seen above for non-disk models with identical initial orientation but otherwise
different parameters such as ARef180 and AExt180.

For halo C the situation is less complicated. The orientation of the gas in the CS09 model, 
which  was used as an initial condition for our model CExtCosmo at $z=1.3$, behaves very similarly to CExtCosmo. It shows an overall reorientation
by the same amount ($\sim 20$ degrees) and comparatively small misalignments, $<45$ degrees.

In figure \ref{align} we compare the orientation of the galaxy (all stars within 10 kpc) 
in a given model to the orientation of the angular momentum of the dark matter (including subhaloes)
within $R$ as a function of radius $R$. Model ARef is initially set up to be aligned with the dark matter within 300 kpc. As can be seen,
this orientation is not, however, aligned with the dark matter within 200 kpc, but the stars and dark matter become aligned by $z=0$ where
the alignment is, in fact, better than 30 degrees for all radii $R<500\;\rm{kpc}$.

In the fully cosmological halo A model the galaxy is aligned with the dark matter within $R\sim 100\;\rm{kpc}$ at $z=1.3$, but is 
strongly misaligned otherwise, especially at $R>150\;\rm{kpc}$, where the angle reaches $\sim 150$ degrees. At $z=0$, there is no disk
and thus the orientation of the galaxy is not well defined. Considering stars within 10 or 30 kpc yields
different results as indicated by the black and blue lines. The stars at $R<10\;\rm{kpc}$ are well-aligned with the central dark halo, whereas
the stars at $10<R/\rm{kpc}<30$ are not. Both populations are misaligned with the dark matter outside $R\sim50\;\rm{kpc}$, but the misalignment
is significantly smaller for the stars at $10<R/\rm{kpc}<30$.

For halo C the picture is different. Model CExtCosmo is initially  misaligned with the dark matter angular momentum by an angle of
$\sim 50$ degrees. This hardly changes by $z=0$, when the inner 100 kpc have become more aligned, but the outer parts remain unchanged.
In the fully cosmological run the situation is strikingly different. 
In the inner $\sim 50\;\rm{kpc}$, the dark matter both at $z=1.3$ and at $z=0.0$ is
strongly misaligned with the galaxy, almost 180 degrees near the centre.
This explains why there is no clear correlation between dark matter and galaxy angular momenta in our models based on halo C.

\vspace{0.5 cm}

In summary, there are no large discrepancies between our models and fully cosmological galaxy formation simulations within the same haloes.
In the absence of any destructive events such as major mergers, a disk will survive over cosmological timescales, if it has formed in the 
preferred orientation for its halo. As we showed, this orientation does not depend on the potential alone (section 5),  nor is there 
a clear correlation with the angular momentum of the halo (this section). The orientations of the halo principal axes and the halo spin axis
can change with time (see \citealp{vera} and \citealp{bett} respectively) and so apparently, can the preferred disk orientation, as is clear from
the halo A run of CS09, where a disk is destroyed by inflowing matter oriented in a new preferred direction. 
Despite this we were able to find initial gas setups which lead to thin disks in this halo.
For halo C there is no such orientation change and thus both our own and the fully cosmological
models produce surviving disks.

\section{Conclusions}

We have presented a series of simulations of idealized SPH models for the formation and evolution of galactic disks within
fully-cosmological $\Lambda$CDM haloes. At $z=1.3$ we add rotating spheres of hot gas in approximate hydrostatic equilibrium to two dark matter
haloes from the Aquarius Project \citep{aquarius}. A parametrized cooling law and standard prescriptions for star
formation allow us to study the evolution of the combined baryon+CDM systems until $z=0$. We study models with different orientations
and amounts of baryonic initial angular momentum, different cooling timescales and different density profiles. This allows us to determine
favorable and non-favorable conditions for the formation and survival of disks. 

Clearly, such simulations are not full models for the formation of disk galaxies. However, they allow
us to study processes relevant to the formation of disks within triaxial and realistically evolving
dark matter haloes, and hence explore the conditions under which massive disks can
exist in $\Lambda$CDM haloes.
This should help us understand the failure or success of various fully cosmological simulations of disk galaxy formation, which still
suffer from a multitude of uncertainties (e.g. \citealp{governato, eris}, CS09, \citealp{cs2011}).
A particular problem may be that these simulations tend to produce overly old stellar populations (see the discussion
in \citealp{ab}) and thus to under-predict the stellar mass formed since $z\sim1$, the epoch generally thought to be
best suited for disk formation due to the absence of major mergers at such late times.
Our models are unable to reproduce full star formation histories
but yield interesting insights into galaxy formation at relatively recent times.

Our setup has an inherent initial discrepancy between the triaxial, substructured and dynamically growing halo
and the rotating quasi-equilibrium gas sphere. This leads to an initial phase, in which the gas adjusts to the
potential. Thereafter it loses angular momentum to the dark matter as it cools 
to the centre and transforms the total central potential, and thus also the halo, into a more
axisymmetric configuration capable of hosting a stable disk. 
We show that the shorter the timescale of this
transformation, the more destructive the effect on the forming disk,
and thus the more dominant the resulting bulge and thick disk components.
Haloes are expected to be near prolate after major mergers
(e.g. \citealp{romanodiaz}) before most of the stellar mass assembles 
into disks, so this process of bulge formation during halo transformation is not unrealistic.
Romano-Diaz et al. also report the initial formation of an asymmetric, bar-unstable disk
in their fully-cosmological simulations.
Only if we align the initial angular momentum of the gas sphere with the major axis of the potential,
can we suppress this destructive process. However, the transformation from a prolate potential
to an oblate disk potential triggers misalignments between inner and outer baryonic components, 
which hinders disk survival.

We have shown that our models are capable of producing, and thus that the haloes in consideration are capable of hosting, disks with 
realistic structural and kinematic properties, which unlike in most cosmological
simulations show realistic disk-to-bulge mass ratios $D/B> 4$. As has been shown previously \citep{cs2010}, structural and kinematic 
decompositions yield strongly differing results, with structural values exceeding kinematic ones by factors of a few. We show that our disks
have rotation-to-dispersion ratios $V_{\rm{rot}}/\sigma_z$ monotonically decreasing with increasing age, which arises from continuous disk
heating due to substructures and a monotonic decrease in birth velocity dispersion connected to the transformation of the central potential.
Due to limitations in resolution, our simulations are not capable of 
producing realistic young thin stellar disks with $\sigma_z< 10 \rm{kms^{-1}}$ 
(see also \citealp{house}), but they agree qualitatively with age-velocity dispersion relations for the solar neighborhood \citep{gcs, ab}.
Our disks show truncations, which move outward with time and the radius of which increases with decreasing formation timescale and increasing 
angular momentum content. 

We find that the slower the formation of a model galaxy, the colder and more dominant is the final disk.
This is the result of the combined effects of the suppression of bulge formation during potential sphericalization
and of continuous disk heating being less efficient for younger populations.
Bars develop in our model disks if their surface density is enhanced by a higher total mass or lower 
total angular momentum. They strongly increase disk heating (see \citealp{saha}), enhance the bulge fraction and produce prominent ring
structures outside the bar regions (see \citealp{athan}) thus lowering the resulting $V_{\rm{rot}}/\sigma_z$ by factors of a few.

We show that the most stable, most extended and coldest disks form from a gas angular momentum
orientation that is aligned with the minor axis of the inner halo, as
had been previously found in cosmological simulations by \citet{bailin}. Out of the two such orientations given by the halo (180 degrees apart)
one is clearly preferred. For one of our two haloes, the preferred direction is determined by the angular momentum of the outer halo in the 
initial conditions; by the end of the simulation all regions are (marginally) aligned with the central disk. The counter-rotating orientation
is strongly disfavored and does not produce stable disks. For the second halo, there is no alignment between halo angular momentum and shape.
The non-preferred minor axis orientation of the gas angular momentum is the one,
which makes the larger angle with the halo spin axis. Models with this initial 
orientation are able to host stable, yet thicker disks.
In each halo we find a certain range of initial gas angular momentum orientations perpendicular to the
halo major axis, for which a reorientation of the baryonic 
spin vector to one of the stable orientations occurs with the disk thickening but surviving. 
Models, where the gas angular momentum is initially aligned with the halo major axis can produce disks, but their orientation is not stable.
Intermediate models cannot produce stable disks.
Reorientation especially heats the outer, lower surface density parts of disks, where self-gravity 
is unable to keep the stellar orbits aligned. Compact, high surface density disks are significantly less heated by orientation changes.

As reorientation is common both for halo shapes \citep{vera} and for halo angular momenta \citep{bett}, the
preferred orientation will evolve with time in $\Lambda$CDM haloes. This is directly connected to the well-known
phenomenon of misaligned matter infall during the assembly of haloes and galaxies (e.g. \citealp{quinn}).
However, in our models, the initial gas distribution is spherically symmetric and its angular momentum is aligned at all radii.
Despite this, the orientations of outer and inner gas and of stellar disks and also
the angle between these components change by more than 90 degrees in many cases.
Misalignment is thus a consequence of strongly nonlinear dynamical interactions in the final stages of galaxy
assembly, not just of evolving asymmetries in the cosmological context, which feeds halo growth.

Misaligned components are common in observed galaxies in the form of warps \citep{sancisi}, polar rings \citep{whitmore},
polar bulges \citep{4698} or counter-rotating components in early-type galaxies \citep{sauron}. Misaligned cosmological infall has 
been identified as a cause for a variety of phenomena, such as the destruction of disks (CS09), the formation of polar 
disks \citep{brook} and the excitation of warps \citep{roskar2}. Interestingly, all of these phenomena also occur in our models.
We qualitatively reproduce the misaligned infall found in the cosmological resimulation of halo A
by CS09 starting from a spherical and coherently rotating gas distribution.
The torques acting in our simulation must originate in the
nonlinear structure of the dark halo, suggesting a mechanism different from the conclusions of \citet{roskar2}
who found negligible influence of the dark halo torques in comparison to anisotropic cosmological infall.

All these processes of misalignment are associated with the loss of angular momentum (see also \citealp{roskar2}) and 
produce more compact, hotter objects. Thin extended disks can thus only form in stable orientations and thus in
haloes for which the preferred disk orientation undergoes little temporal evolution.
This conclusion is similar to that of \citet{sales}, who found that disks form from the continuous accretion of material
with coherently aligned angular momentum orientation and that misaligned accretion of gas tends to produce spheroids.
The question arises whether $\Lambda$CDM offers enough haloes which fulfill these conditions to explain the
observed abundance of disk galaxies in the universe. \citet{sales} find a significant fraction of such disk-dominated
systems in their model sample of galaxies, but the limitations of their simulations do not allow a full comparison
with the statistical properties of observed samples.
Semi-analytic models of galaxy formation (e.g. \citealp{guo2}) are helpful for studying galaxy populations, but they do not
capture the role of misalignments and other dynamical details that, as we have shown, play a major role in galaxy evolution.
It would be desirable to add proper treatment of these processes to the models, however, we are not able to offer straightforward
prescriptions, as we are not capable of drawing general conclusions from the presented study of two haloes.

Considering our work and recent improvements in cosmological resimulations of individual galaxies (e.g. \citealp{eris, 4disk}),
realistic $\Lambda$CDM haloes appear to be capable of hosting realistic disk galaxies.
However, more work is needed to understand whether the population of $\Lambda$CDM 
haloes agrees in detail with the observed population of galaxies.

\section*{Acknowledgments}

We are grateful to Jackson DeBuhr, Chung-Pei Ma, Laura Sales and Ralph Sch{\"o}nrich for valuable discussions.
We thank Cecilia Scannapieco for kindly providing and discussing her cosmological simulations.
MA acknowledges support from the DFG Excellence Cluster "Origin and Structure of the Universe".

\end{document}